\documentclass[%
 reprint,
 amsmath,amssymb,
 aps,
]{revtex4-1}
\usepackage[latin1]{inputenc}
\usepackage{mathtools}
\usepackage{slashed}
\usepackage{physics}
\usepackage{array}
\usepackage{amsmath}
\usepackage[dvipsnames]{xcolor}
\usepackage{amsfonts}
\usepackage{amssymb}
\usepackage{relsize}
\usepackage{graphicx}
\usepackage{amssymb,epsfig,subfigure}
\usepackage{hyperref}
\usepackage{comment}
\usepackage{tabularx}
\usepackage{bm}
\usepackage{euscript}
\usepackage{graphicx}
\usepackage{mathrsfs}
\usepackage[scr=boondox]{mathalfa}
\usepackage{color}
\usepackage [english]{babel}
\usepackage [autostyle, english = american]{csquotes}
\MakeOuterQuote{"}
\usepackage{amsfonts}
\usepackage{exscale}
\usepackage{amsbsy}
\usepackage{subfigure}
\usepackage{textcomp}
\usepackage{comment}
\usepackage{hyperref}
\usepackage{slashed}
\usepackage{tabularx}
\usepackage{euscript}
\usepackage{graphicx}
\usepackage{color}
\usepackage{amsfonts}
\usepackage{exscale}
\usepackage{amsbsy}
\usepackage{subfigure}
\usepackage{textcomp}
\usepackage{comment}
\usepackage{hyperref}
\usepackage{bm}
\usepackage{wrapfig}
\usepackage{ulem}
\usepackage{lipsum}

\usepackage[font=footnotesize,labelfont=bf]{caption}

\newcommand{\beq}{\begin{eqnarray}}
\newcommand{\eeq}{\end{eqnarray}}
\begin{document}

\def\ppnumber{\vbox{\baselineskip14pt
}}

\def\ppdate{
} \date{\today}

\title{\bf Interaction Enabled Fractonic Higher-Order Topological Phases}
\author{Julian May-Mann}
\affiliation{Department of Physics, University of Illinois Urbana-Champaign, Urbana,IL 61801, USA}
  \author{Yizhi You}
  \affiliation{Department of Physics, University of Oxford, Oxford, OX1 3PJ, UK}
  \author{Taylor L. Hughes}
  \affiliation{Institute for Condensed Matter Theory, University of Illinois Urbana-Champaign, Urbana,IL 61801, USA}
\author{Zhen Bi}
\affiliation{Department of Physics, The Pennsylvania State University, University Park, Pennsylvania 16802, USA} 
\date{\today}

\begin{abstract}
 In this work, we present a collection of three-dimensional higher-order symmetry protected topological phases (HOSPTs) with gapless hinge modes that exist only in strongly interacting systems subject to subsystem symmetry constraints. We use a coupled wire construction to generate three families of microscopic lattice models: insulators with helical hinge modes,  superconductors with chiral Majorana hinge modes, and fractionalized insulators with helical hinge modes that carry fractional charge. In particular, these HOSPTs do not require spatial symmetry protection, but are instead protected by subsystem symmetries, and support "fractonic" quasiparticle excitations that move within only a low-dimensional sub-manifold of the system. We analyze the anomaly structure for the boundary theory and the entanglement Hamiltonian, and show that the side surfaces of these HOSPTs, despite being partially gapped, exhibit symmetry anomalies, and can only be realized as the boundary of three-dimensional HOSPT phases.
\end{abstract}

\maketitle

\bigskip
\newpage

\section{Introduction}
Higher-order symmetry protected topological phases (HOSPTs) are novel forms of gapped quantum matter that host gapped surfaces but, nonetheless, have gapless hinge or corner modes protected by symmetry\cite{benalcazar2017quantized,schindler2018higher}. Since their initial discovery, HOSPTs have attracted a great deal of attention from both the theoretical and experimental communities. Recent progress includes symmetry classifications\cite{benalcazar2017electric,song2017d,langbehn2017reflection,khalaf2018higher,benalcazar2019quantization,you2021multipolar}, topological field theory descriptions\cite{you2021multipolar, may2021crystalline,you2018higher,you2019higher}, and experimental realizations of various classes of HOSPTs\cite{noh2018topological,serra2018observation,peterson2018quantized,imhof2018topolectrical,schindler2018higherb,xue2019acoustic,zhang2019second,ni2019observation,noguchi2021evidence,aggarwal2021evidence}. Despite the success in non-interacting HOSPTs, i.e., band insulators and mean-field superconductors, the study of strongly interacting HOSPTs is still in an early stage.  

Introducing unconventional symmetries has proven to be a sucessful way to expand the family of topological phases of matter. As an intriguing generalization of conventional global symmetry, subsystem symmetry is a symmetry that acts independently along sub-regions of the whole physical system. Previous works have established that subsystem symmetry can restrict the mobility of charged excitations in fracton phases of matter\cite{vijay2016fracton, nandkishore2019fractons, pretko2020fracton,pai2019fracton}, and can lead to a number of new topological phases\cite{you2018subsystem,devakul2019fractal,you2018subsystem,devakul2018classification,devakul2020strong,williamson2019fractonic,may2019corner,you2020symmetric, devakul2020strong,stephen2020subsystem,may2021topological}. For instance, based on field theoretical considerations, Ref.~ \cite{you2021multipolar} proposed a new type of higher-order topological insulator that has chiral hinge modes protected by two distinct subsystem symmetries which act along $2$D $xz$ and $yz$-planes respectively. In addition to the interesting hinge modes, the $2$D planar symmetry forbids single-charge hopping terms in the $x$ and $y$-directions, therefore, individual charge excitations are sub-dimensional in the bulk and can only move along the $z$-direction.

In this article, motivated by recent developments in subsystem symmetric topological phases\cite{you2021multipolar,you2019higher}, we construct microscopic lattice models of 3D subsystem symmetric HOSPTs that host gapless hinge modes. Interestingly, the HOSPTs protected by the subsystem symmetries do not require spatial symmetries to protect their hinge modes. This is in stark contrast to the vast majority of previously studied HOSPTs where spatial symmetries are necessary for the stability of the corner or hinge modes. More importantly, because of the subsystem symmetries, single particle tunneling along certain directions is forbidden. Therefore, any models that respect these symmetries, and have interesting dynamics, are intrinsically strongly interacting and have no non-interacting counterparts with similar underlying physics. 

Motivated by these unusual features, we seek to better understand the class of subsystem-symmetric HOSPTs by constructing and analyzing three illustrative examples: a topological insulator having helical fermion hinge modes that are protected by U$(1)$ subsystem symmetries and a $\mathbb{Z}_2$ global symmetry, a topological superconductor having chiral Majorana hinge modes that are protected by $\mathbb{Z}_2$ subsystem symmetries, and a fractional topological insulator having fractionalized helical hinge modes that are protected by U$(1)$ subsystem symmetries in additional to a $\mathbb{Z}_2$ global symmetry. To develop these models we employ various coupled wire constructions\cite{poilblanc1987quantized, teo2014luttinger,vazifeh2013weyl, meng2015fractional, sagi2015array, iadecola2016wire, zhang2022coupledwire} to explicitly construct strongly interacting subsystem symmetric insulators/superconductors with fully gapped bulk and side surfaces, and gapless hinges supporting $1$D chiral or helical modes. We show that these $1$D modes are anomalous and can exist \textit{only} as hinge modes of a $3$D topological phase. We point out that this is rather peculiar, as the conventional expectation is that an anomalous $N$-dimensional system with an onsite symmetry (i.e., not a subsystem or spatial symmetry) can be realized on the surface of a symmetric $N+1$-dimensional bulk\cite{callan1985anomalies,chen2011two,senthil2015symmetry}. Here, however, the $1$D hinge anomalies associated with subsystem symmetry require a $3$D bulk (i.e., an $N+2$-dimensional bulk), and cannot be realized as the edge modes of any $2$D lattice model having the same subsystem symmetry. We expect that this anomaly argument can be generalized to subsystem symmetries in higher dimensions as well.

The remainder of the paper is structured as follows. In Sec. \ref{sec:HOTI} we present and analyze a subsystem symmetric insulator with helical hinge modes. In Sec. \ref{sec:HOTSC} we present and analyze a subsystem symmetric superconductor with four chiral Majorana modes on each hinge. In Sec. \ref{sec:FHOTI} we present and analyze a subsystem symmetric insulator having fractionalized helical hinge modes. For each of these models, we explicitly show that the bulk and boundaries are gapped by symmetric interactions, and that the hinges are gapless and anomalous with respect to the subsystem symmetry. We also include several appendices that contain related constructions and technical details. 

It is also worth mentioning that several $3$D symmetry enriched fractonic phases have recently been constructed using the coupled wire formalism\cite{you2020symmetric,shirley2019foliated,sullivan2020fractonal,sullivan2021fractonic,burnell2021anomaly}. Similar to the subsystem symmetric HOSPTs we present here, these fracton models have sub-dimensionally mobile excitations, but unlike the HOSPTs, these models have fully dispersive gapless surface modes instead of gapped surfaces and gapless hinge modes.

\section{Higher order topological insulator with subsystem charge conservation}
\label{sec:HOTI}
In this section, we provide a microscopic construction of a spin-1/2 fermionic HOSPT that is protected by subsystem U$(1)$ symmetry and a global $\mathbb{Z}_2$ symmetry. The U$(1)$ subsystem symmetry corresponds to the conservation of charge along each $xz$ and $yz$-plane, and the $\mathbb{Z}_2$ symmetry corresponds to a global conservation of spin parity. 
We will find that this strongly-correlated model will exhibit a gapped bulk and surface, but will harbor helical hinge modes. 
\subsection{Fermionic Wire Model}
\begin{figure*}
\includegraphics[width=0.3\textwidth]{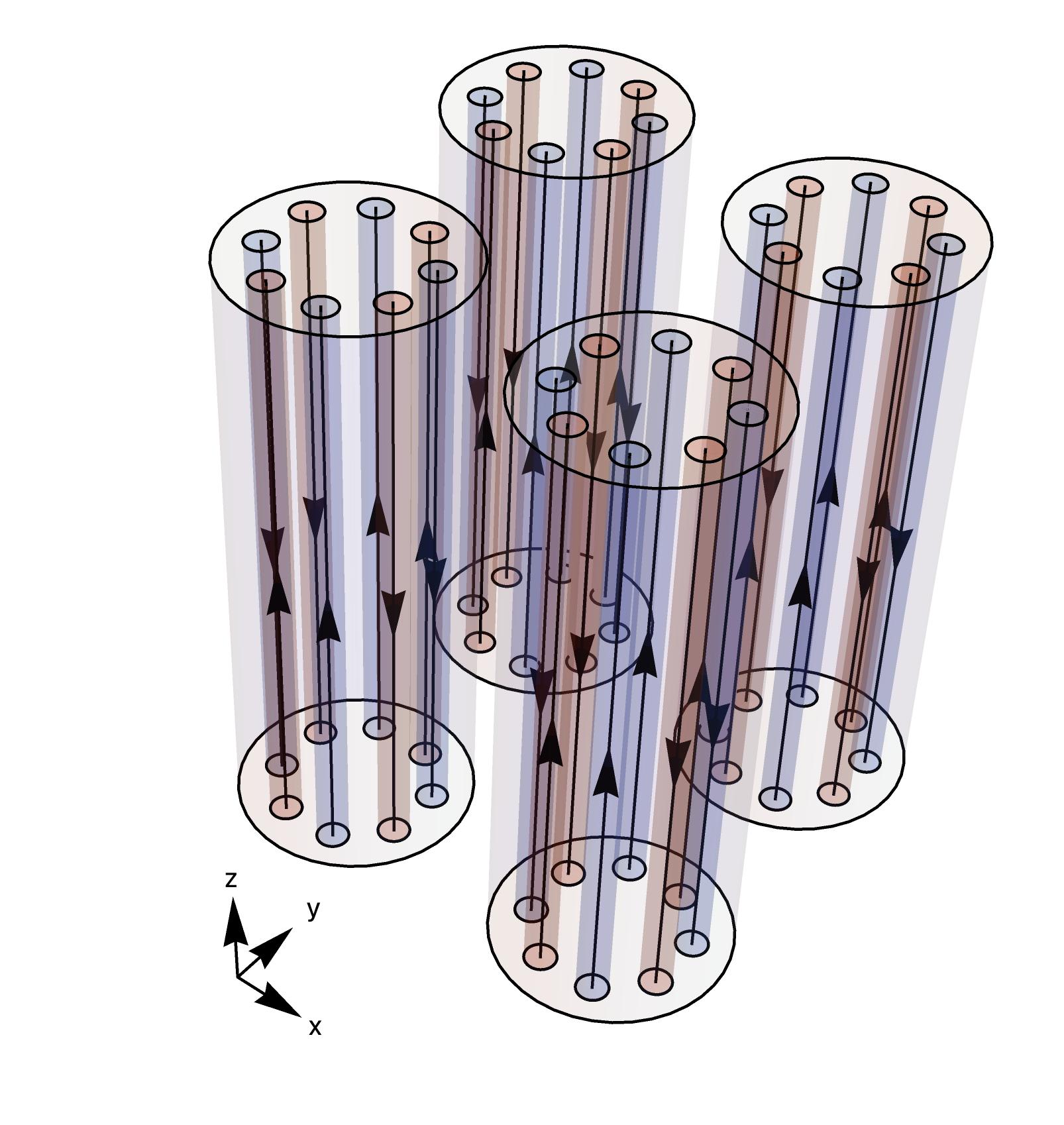}
\includegraphics[trim=0 -0.5in 0 0, clip, width=0.28\textwidth]{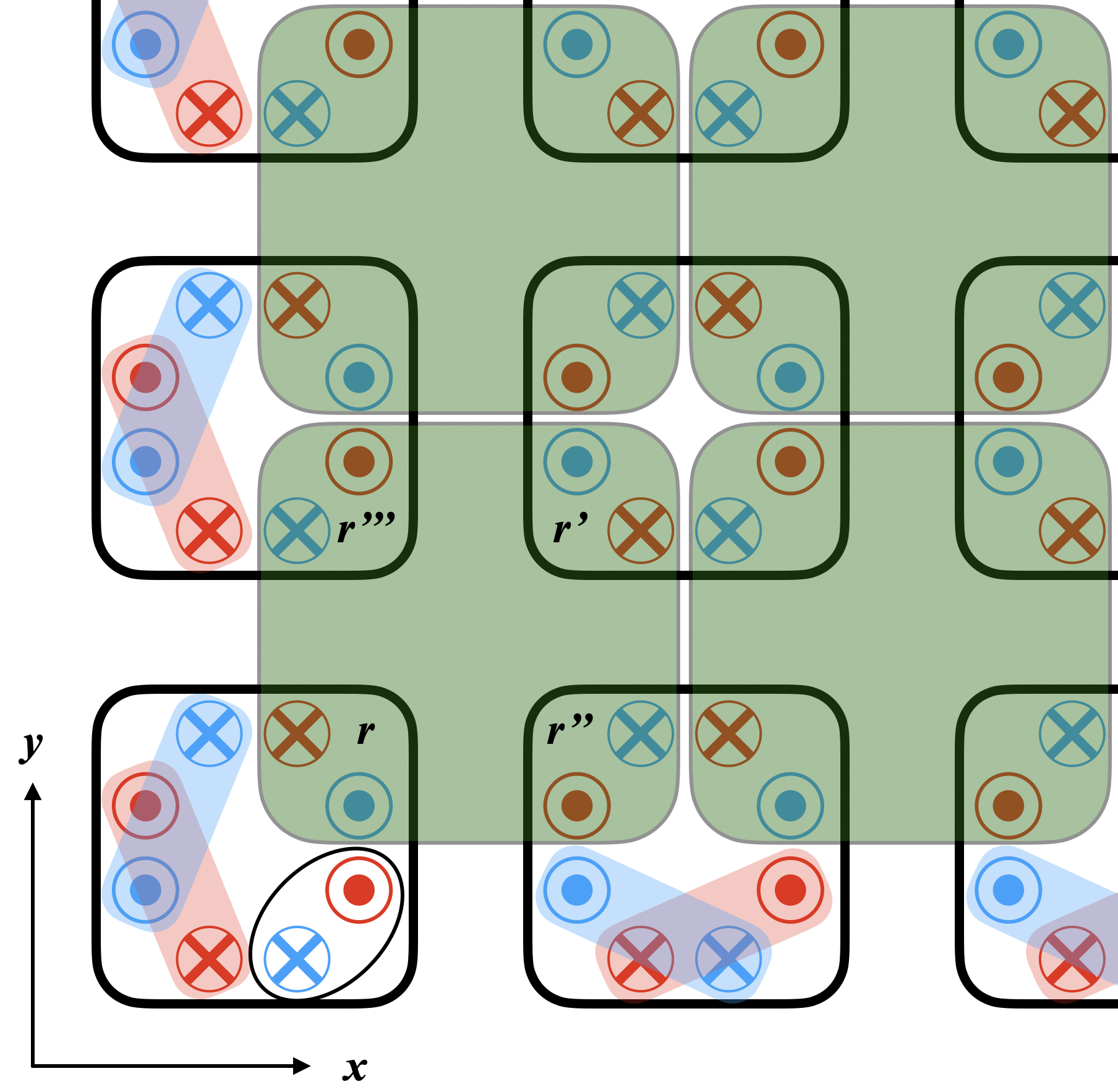}
\includegraphics[trim=-1.5in -0.5in 0 0, clip, width=0.32\textwidth]{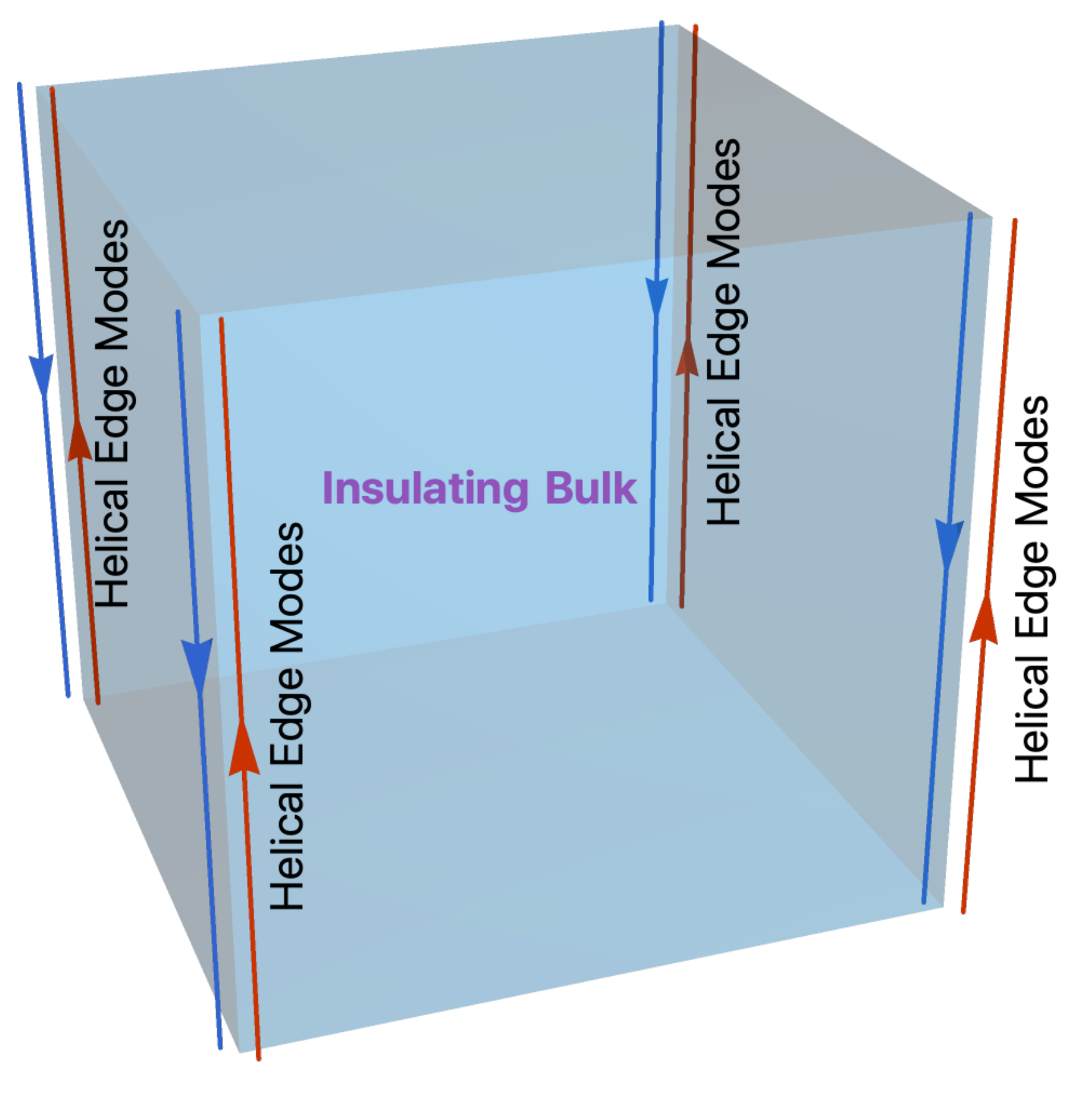}
\caption{(Left) the array of wires used to construct the HOTI. (Middle) the projected view in the $xy$-plane. Each unit cell (denoted by the black square) is composed of eight fermion modes, four upward chiral modes ($\cdot$), and four downward chiral modes, ($\times$). The blue/red color denotes the $\uparrow$/$\downarrow$ fermions that carry opposite global $Z_2$ charges. The green blocks denote the inter-unit-cell interaction which gaps out the fermion modes within the block. On the surface of the system, there are symmetry allowed intra-unit-cell tunnelings (denoted by the blue and red blocks in the surface unit cells) which gap out the surface modes. However, there are symmetry protected hinge modes when two perpendicular surfaces intersect. (Right) A schematic picture for the higher order topological insulator. 
}
\label{fig:basis}
\end{figure*}

Let us consider a $2$D lattice (which spans the $xy$ plane) of $1$D wires (which span the $z$-direction). To be explicit, we will use a square lattice in the $xy$-plane with four $1$D Dirac fermions per unit cell, two of which are spin up and two of which are spin down as in Fig. \ref{fig:basis}. For simplicity, we assume that the Dirac fermions correspond to excitations which are all at the same lattice momentum.   The low energy Hamiltonian for the wires is 
\begin{equation}
    \mathcal{H}_{\text{wires}}=\sum_{\bm{r}} \bm{\psi}^\dagger_{\bm{r}} i\partial_z \tau^{zz0}\bm{\psi}_{\bm{r}},
    \end{equation} 
where  $\bm{r} = n_x \hat{x} + n_y \hat{y}$ labels the unit cells of the $2$D lattice, and $\bm{\psi} = (\psi^1_{\uparrow L}, \psi^2_{\uparrow L}, \psi^3_{\uparrow R}, \psi^4_{\uparrow R}, \psi^1_{\downarrow R}, \psi^2_{\downarrow R}, \psi^3_{\downarrow L}, \psi^4_{\downarrow L})$. The $R$ and $L$ subscript indicates that the mode propagates along the $+z$ or $-z$-direction, respectively. Here, and throughout, we use the shorthand, $\tau^{ij...k} \equiv \tau^i \otimes \tau^j \otimes...\otimes \tau^k$, where $\tau^i$ are the Pauli matrices, including the identity. Hence, the first index of $\tau^{ijk}$ acts on the spin index, while the other two indices act on the flavors of a given spin.

Now we can choose couplings between the wires to gap out the bulk. To generate something topological, we expect that we will need couplings between unit cells. The wires can be gapped with single-particle inter-cell tunneling terms, however such couplings between unit cells would violate subsystem symmetries. Instead, we will couple the wires together using quartic, subsystem-symmetric inter-cell interactions. The interactions are specifically chosen such that the model decomposes into a decoupled array of $1$D wire bundles when viewing from the $xy$ plane. These $1$D bundles are composed of fermionic modes from the lattice sites that surround a given plaquette, as shown in see Fig. \ref{fig:basis}.  The HOTI Hamiltonian we consider is
\begin{equation}\begin{split}
    &\mathcal{H}_{\text{HOTI}}= \mathcal{H}_{\text{wires}} + \mathcal{H}_{\text{int}},\\
    &\mathcal{H}_{\text{int}} = \sum_{\bm{r}} \Big[V_1 \psi^{1 \dagger}_{\uparrow L, \bm{r}}\psi^{2\dagger}_{\uparrow L,\bm{r}'}\psi^3_{\uparrow R,\bm{r}''}\psi^4_{\uparrow R,\bm{r}'''}\\
    &\phantom{=======} + V_2 \psi^{1 \dagger}_{\downarrow R, \bm{r}}\psi^{2\dagger}_{\downarrow R,\bm{r}'}\psi^3_{\downarrow L,\bm{r}''}\psi^4_{\downarrow L,\bm{r}'''}\\
    &\phantom{=======} + V_3 \psi^{1 \dagger}_{\uparrow L, \bm{r}}\psi^3_{\uparrow R,\bm{r}''}\psi^{ 1}_{\downarrow R, \bm{r}}\psi^{3\dagger}_{\downarrow L,\bm{r}''}\\
    &\phantom{=======} + V_4 \psi^{1 \dagger}_{\uparrow L, \bm{r}}\psi^4_{\uparrow R,\bm{r}'''}\psi^{ 1}_{\downarrow R, \bm{r}}\psi^{4\dagger}_{\downarrow L,\bm{r}'''}\Big] + h.c. \\
\label{eq:LatticeHamiltonianHOTI}\end{split}\end{equation}
where  $\bm{r} = n_x \hat{x} + n_y \hat{y},$ $\bm{r}' \equiv \bm{r}+\hat{x}+\hat{y}$, $\bm{r}'' \equiv \bm{r}+\hat{x}$ and $\bm{r}''' \equiv \bm{r}+\hat{y}$. 
$\mathcal{H}_{\text{HOTI}}$ has a global $\mathbb{Z}_2$ symmetry,
\begin{equation}\begin{split}
\mathbb{Z}_2: \bm{\psi} \rightarrow \tau^{z00}\bm{\psi},
\label{eq:Z2Def}\end{split}\end{equation}
which corresponds to the global conservation of spin parity. Additionally, it has two U$(1)$ subsystem symmetries,
\begin{equation}\begin{split}
\text{U}(1)^{xz}: \bm{\psi}_{\bm{r}} \rightarrow e^{i \theta_{xz}(\bm{r}\cdot\hat{y})} \bm{\psi}_{\bm{r}},\\
\text{U}(1)^{yz}: \bm{\psi}_{\bm{r}} \rightarrow e^{i \theta_{yz}(\bm{r}\cdot\hat{x})}\bm{\psi}_{\bm{r}},
\label{eq:SubSystemDefU(1)}\end{split}\end{equation}
where $\theta_{xz}$ is a real function of $\bm{r}\cdot \hat{y} = n_y$, and $\theta_{xy}$ is a real function of $\bm{r}\cdot \hat{x} = n_x$. These subsystem symmetries indicate that charge is conserved along both $xz$-planes and $yz$-planes (and by extension the total charge is conserved as well). As mentioned above, we see that single particle inter-cell tunneling in either the $x$ or $y$-directions necessarily breaks these symmetries, although tunneling along the $z$-direction (i.e., along the wires) is allowed.

From our choice of interactions, the fermions $\psi^{ 1}_{\uparrow L, \bm{r}}$, $\psi^{2}_{\uparrow L,\bm{r}'}$, $\psi^3_{\uparrow R,\bm{r}''}$, $\psi^4_{\uparrow R,\bm{r}'''}$, $\psi^{ 1}_{\downarrow R, \bm{r}}$, $\psi^{2}_{\downarrow R,\bm{r}'}$, $\psi^3_{\downarrow L,\bm{r}''}$, and $\psi^4_{\downarrow L,\bm{r}'''}$ only couple to one another, for a fixed $\bm{r}$. This set of fermions can be treated as a $1$D wire bundle, and the full $3$D model is simply a 2D array of these bundles.  If each bundle is fully gapped (see below for an explicit calculation), then the bulk of the HOTI is also fully gapped. Furthermore, if the system have open surfaces perpendicular to $x$ or $y$ direction, then each surface unit cell will be left with a pair of helical modes, which can be gapped out with an intra-unit-cell tunneling that preserves all the symmetries. However, on the hinges there will be an odd number of helical modes. For example, for hinges between boundaries normal to the $+x$ and $+y$-directions the fermions $\psi^1_{\uparrow L}$, $\psi^1_{\downarrow R}$, $\psi^3_{\uparrow R}$, $\psi^3_{\downarrow L}$, $\psi^4_{\uparrow R}$, and $\psi^4_{\downarrow L}$ are gapless, and there is a net positive helicity mode. This net helical mode is protected from acquiring a gap by the global $\mathbb{Z}_2$ symmetry. There are similar helical modes on the other hinges, as well.

To show that the interactions $\mathcal{H}_{\text{int}}$ in Eq. \ref{eq:LatticeHamiltonianHOTI} fully gap the bulk of our system, we will use bosonization. We identify the fermionic operators with the vertex operators $\psi^j_{\sigma R/L} \sim e^{\mp i \phi^1_{\sigma R/L}}$, where $\sigma = \uparrow,\downarrow,$ and the $\mp$ are correlated with  the $R/L$ subscript. In terms of these bosonic variables, the Lagrangian corresponding to Eq. \ref{eq:LatticeHamiltonianHOTI} is
\begin{equation}\begin{split}
\mathcal{L} = &\sum_{\bm{r}} \Big[ -\partial_t \bm{\phi}_{\bm{r}}^T\tau^{zz0} \partial_z \bm{\phi}_{\bm{r}} - \partial_z \bm{\phi}_{\bm{r}}^T V  \partial_z \bm{\phi}_{\bm{r}}\\ & -g_1 \cos(\phi^1_{\uparrow L,\bm{r}} + \phi^2_{\uparrow L,\bm{r}'} + \phi^3_{\uparrow R,\bm{r}''} + \phi^4_{\uparrow R,\bm{r}'''})  \\ & -g_2 \cos(\phi^1_{\downarrow R,\bm{r}} + \phi^2_{\downarrow R,\bm{r}'} + \phi^3_{\downarrow L,\bm{r}''} + \phi^4_{\downarrow L,\bm{r}'''}) \\ & -g_3 \cos(\phi^1_{\uparrow L,\bm{r}} + \phi^3_{\uparrow R,\bm{r}''} -\phi^1_{\downarrow R,\bm{r}}-\phi^3_{\downarrow L,\bm{r}''})  \\ & -g_4 \cos(\phi^1_{\uparrow L,\bm{r}} + \phi^4_{\uparrow R,\bm{r}'''}- \phi^1_{\downarrow R,\bm{r}}- \phi^4_{\downarrow L,\bm{r}'''})\Big],
\label{eq:BosonLagrangianHOTI}\end{split}\end{equation} where $\bm{\phi} = (\phi^1_{\uparrow L},\phi^2_{\uparrow L},\phi^3_{\uparrow R},\phi^4_{\uparrow R},\phi^1_{\downarrow R},\phi^2_{\downarrow R},\phi^3_{\downarrow L},\phi^4_{\downarrow L})$ and $V$ is an $8 \times 8$ "velocity" matrix, which includes the fermionic kinetic energy terms as well as various forward-scattering terms. Crucially, all the cosine terms in Eq. \ref{eq:BosonLagrangianHOTI} commute with one another. Hence, provided that the $g_i$ couplings are significantly strong, the bosonic fields $\phi^{ 1}_{\uparrow L, \bm{r}}$, $\phi^{2}_{\uparrow L,\bm{r}'}$, $\phi^3_{\uparrow R,\bm{r}''}$, $\phi^4_{\uparrow R,\bm{r}'''}$, $\phi^{ 1}_{\downarrow R, \bm{r}}$, $\phi^{2}_{\downarrow R,\bm{r}'}$ $\phi^3_{\downarrow L,\bm{r}''}$, and $\phi^4_{\downarrow L,\bm{r}'''}$ are massive for each wire bundle labeled by $\bm{r}$. As noted before, this fully gaps the bulk, while leaving behind gappable surface modes, and symmetry protected hinge modes. Based on this, we conclude that at strong coupling the fermionic model in Eq. \ref{eq:LatticeHamiltonianHOTI} realizes a higher-order topological phase with protected with helical hinge modes.

\subsection{O$(4)$ Non-Linear Sigma Model}\label{ssec:NLSM}
It will be instructive to provide an alternative perspective of our wire construction using well-known properties of $1$D non-linear sigma models.
As noted before, the fermions in Eq. \ref{eq:LatticeHamiltonianHOTI} form decoupled wire bundles of four gapless Dirac fermions. To study the properties of a single bundle, let us define $\tilde{\psi}_{\uparrow \bm{r}} = (\psi^{ 1}_{\uparrow L, \bm{r}},\psi^{2}_{\uparrow L,\bm{r}'},\psi^3_{\uparrow R,\bm{r}''},\psi^4_{\uparrow R,\bm{r}'''})$ and  $\tilde{\psi}_{\downarrow\bm{r}} = (\psi^{ 1}_{\downarrow R, \bm{r}},\psi^{2}_{\downarrow R,\bm{r}'},\psi^3_{\downarrow L,\bm{r}''},\psi^4_{\downarrow L,\bm{r}'''}).$ The global $\mathbb{Z}_2$ symmetry acts as $\tilde{\psi}_{\uparrow \bm{r}} \rightarrow \tilde{\psi}_{\uparrow \bm{r}}$ and $\tilde{\psi}_{\downarrow \bm{r}} \rightarrow -\tilde{\psi}_{\downarrow \bm{r}}$ on these degrees of freedom. Our Hamiltonian is such that the fermions $\tilde{\psi}_{\uparrow \bm{r}}$ and $\tilde{\psi}_{\downarrow \bm{r}}$ only couple to one another for a fixed $\bm{r}$.  

To understand the underlying physics of the subsystem symmetric HOTI, let us first consider only the four fermions $\tilde{\psi}_{\uparrow\bm{r}}$ with fixed $\bm{r}$ (we will drop the subscript $\bm{r}$ for the rest of the subsection for brevity). Taking the linear combinations of the subsystem symmetry defined in Eq. \ref{eq:SubSystemDefU(1)}, we can define 3 independent U(1) symmetries that act as following
\begin{equation}\begin{split}
&\tilde{\text{U}}(1)^{xz}: \tilde{\psi}_\uparrow\rightarrow e^{i\tilde{\theta}_{xz}\tau^{zz}}\tilde{\psi}_\uparrow,\\ 
&\tilde{\text{U}}(1)^{yz}: \tilde{\psi}_\uparrow\rightarrow e^{i\tilde{\theta}_{yz}\tau^{0z}}\tilde{\psi}_\uparrow,\\ 
&\text{U}(1)^{total}: \tilde{\psi}_\uparrow\rightarrow e^{i\theta \tau^{00}}\tilde{\psi}_\uparrow.
\label{eq:U1ss}
\end{split}\end{equation} 
We find that these symmetry assignments are very similar to those of the edge states of two copies of a $2$D quantum spin hall (QSH) insulator exhibiting both total charge and spin $S_z$ conservation, i.e., $\text{U}(1)_c\times \text{U}(1)_s$ symmetry (not to be confused with the charge and spin of the HOTI in Eq. \ref{eq:LatticeHamiltonianHOTI}).  More explicitly, we can map $\tilde{\text{U}}(1)^{xz}_\uparrow \rightarrow \text{U}(1)_c$ and $\tilde{\text{U}}(1)^{yz}_\uparrow \rightarrow \text{U}(1)_s$. The interacting classification of $2$D QSH with $\text{U}(1)_c\times \text{U}(1)_s$ is labeled by an integer $\nu$\cite{bi2015classification}, and the $\tilde{\psi}_\uparrow$ fermions correspond to the edge states of a $\nu=2$ system. Therefore, interactions cannot fully gap the $\tilde{\psi}_\uparrow$ fermions due to the anomaly associated with the two $\text{U}(1)$ symmetries (though, as we have seen, it {\it{is}} possible to fully gap the $\tilde{\psi}_\uparrow$ fermions by coupling them to the $\tilde{\psi}_\downarrow$ fermions). 

Although we see that interactions cannot fully gap the spectrum of the  $\tilde{\psi}_\uparrow$ fermions, it can gap out all single-particle fermion modes such that the spectrum consists of only gapless bosonic modes\cite{bi2015classification,you2015bridging}. These gapless bosonic degrees of freedom can be described by an effective field theory of an $\text{O}(4)$ nonlinear sigma model with Wess-Zumino-Witten (WZW) term at level-1 (possibly with anisotropic terms). 

To see this, we couple the $\tilde{\psi}_{\uparrow}$ fermion to a fluctuating O$(4)$ order parameter $\vec{m}_\uparrow = (m_{1\uparrow},m_{2\uparrow},m_{3\uparrow},m_{4\uparrow})$:
\begin{equation}\begin{split}
H_{\text{O}(4),\uparrow}=\tilde{\psi}^{\dagger}_{\uparrow}\Big[&i\partial_z \tau^{z0}+m_{1\uparrow}\tau^{xx}+m_{2\uparrow}\tau^{xy}\\
&+m_{3\uparrow}\tau^{xz}+m_{4\uparrow} \tau^{y0}\Big]\tilde{\psi}_\uparrow.
\label{eq:hinge}\end{split}\end{equation}
The bilnear fermion terms should be considered as the result of a Hubbard-Stratonovich transformation of quartic fermionic interactions $V_1 \psi^{1 \dagger}_{\uparrow L, \bm{r}}\psi^{2\dagger}_{\uparrow L,\bm{r}'}\psi^3_{\uparrow R,\bm{r}''}\psi^4_{\uparrow R,\bm{r}'''}$ in Eq. \ref{eq:LatticeHamiltonianHOTI}. The $\tilde{\psi}^{\dagger}_{\uparrow}$ fermions are massive due to the O$(4)$ order parameter, and can be integrated out\cite{abanov2000}. The resulting effective theory for the O$(4)$ order parameter is a non-linear sigma model including a WZW term:
\begin{equation}\begin{split}
&\mathcal{L}_\uparrow=\frac{1}{g}(\partial_{\mu} \vec{m}_\uparrow)^2+\frac{2\pi}{\Omega^3} \int_0^1 du\epsilon^{ijkl}  m_{i\uparrow}\partial_z m_{j\uparrow} \partial_t m_{k\uparrow}\partial_u m_{l\uparrow},\\
 &\vec{m}_\uparrow(x,t,u=0)=(1,0,0,0),\phantom{=}
 \vec{m}_\uparrow(x,t,u=1)=\vec{m}_\uparrow(x,t),
\end{split}
\label{eq:NLSM1}\end{equation}
which represents an $SU(2)_1$ conformal field theory in 1+1D.

For the bosonic degrees of freedom we find that the subsystem symmetry $\tilde{\text{U}}(1)^{xz}$ rotates $m_{3\uparrow}$, and $m_{4\uparrow}$ by $\tilde{\theta}_{xz}$ while $\tilde{\text{U}}(1)^{yz}$ rotates $m_{1\uparrow}$, and $m_{2\uparrow}$ by $\tilde{\theta}_{yz}$. Thus, the WZW term implies a perturbative anomaly\cite{xu2013wave,bi2015classification} of the $\tilde{\text{U}}(1)^{xz}\times\tilde{\text{U}}(1)^{yz}$ symmetry, whose physical effect is that a $2\pi$ flux insertion for $\tilde{\text{U}}(1)^{xz}$ would carry a unit charge of $\tilde{\text{U}}(1)^{yz}$. Hence, due to the perturbative anomaly, the spectrum of the spin-up half of the wire bundle is robustly gapless against any perturbation provided the symmetry is not broken.

Now let us reintroduce the $\tilde{\psi}_\downarrow$ fermions. Following the same logic as before, the low energy effective field theory of these fermions will be an O$(4)$ WZW theory, but at level  $k=-1$ instead of $k=+1$ since its kinetic energy term has the opposite sign of $\tilde{\psi}_\uparrow$. When combined together we obtain
\begin{equation}\begin{split}
\mathcal{L}=&\sum_{\sigma} \frac{1}{g}(\partial_{\mu} \vec{m}_\sigma)^2 \\ &+ \frac{(-1)^{\sigma} 2\pi}{\Omega^3} \int_0^1 du\epsilon^{ijkl}  m_{i\sigma}\partial_z m_{j\sigma} \partial_t m_{k\sigma}\partial_u m_{l\sigma}.
\end{split}\end{equation}
The anomaly associated with the $\tilde{\text{U}}(1)^{xz}\times\tilde{\text{U}}(1)^{yz}$ symmetry cancels in the combined theory. In addition, the global $\mathbb{Z}_2$ symmetry acts trivially on the O$(4)$ order parameters. Therefore, there is no anomaly reason to prevent us from gapping out the combined system. As an example, we can turn on a coupling $-A(m_{1\uparrow}m_{1\downarrow}+m_{2\uparrow}m_{2\downarrow}+m_{3\uparrow}m_{3\downarrow}+m_{4\uparrow}m_{4\downarrow})$, which preserve the $\tilde{\text{U}}(1)^{xz}\times\tilde{\text{U}}(1)^{yz}$ symmetry. In the large $A>0$ limit, the system will energetically prefer the field configuration where $m_{i\uparrow}=m_{i\downarrow}$ for all the components. In this limit, the WZW terms cancel each other, and we get a pure O$(4)$ nonlinear sigma model in $1$D which will flow to a gapped symmetric phase at low energy. This gapped bulk is generated by a dynamical mass due to the strong interactions, and the resulting state does not break any symmetry of the system. This phenomenon is sometimes refereed as dynamical mass generation in the literature\cite{you2015bridging,he2018dynamical}.

\subsection{Hinge Anomalies}\label{ssec:HingeAnom}

From our construction we have shown there exists a subsystem symmetric 3D topological insulator that supports helical hinge modes. However, having gapless modes at the surface (or hinge) does not immediately guarantee that the bulk is topological.
To further reveal the bulk topological nature of our model, we will now demonstrate that our helical hinge modes cannot exist as the edge of a purely 2D lattice model with local symmetric interactions. To elucidate this, we begin with the HOTI model in Eq. \ref{eq:LatticeHamiltonianHOTI} and take PBCs along the $z$-direction. Leaving the other directions open, we see that the boundary contains the four side surfaces on the $xz/yz$ planes with four helical hinge states along the four hinges. We now aim to demonstrate the anomalous nature of this `boundary' by  providing a no-go theorem that a similar boundary state cannot be realized on a pure 2D `cover' with the same symmetry assignment.

To begin, let us take a lattice model placed on a 2D $xz$-plane. We assume this model carries a global $\mathbb{Z}_2$ symmetry as well as the subsystem U(1) symmetry, where the charge is conserved on each $z$-row. If we are able to construct a model that: (i) respects these symmetries, (ii) has a gapped bulk, and (iii) has a gapless boundary harboring helical modes carrying opposite $\mathbb{Z}_2$ charge propagating along the $z$-edge, then one can attach such $2$D sheets on two opposing side surfaces of the HOTI and the helical hinge states can be eliminated. If this were possible, then the helical hinge state would not be a signature of the bulk topology, since they could be destroyed by surface reconstruction while preserving symmetry. However, we will argue that such a 2D sheet would exhibit a global anomaly, and hence cannot be realized in a lattice model with local interactions. 

\begin{figure}
\centering
\includegraphics[width=0.09\textwidth]{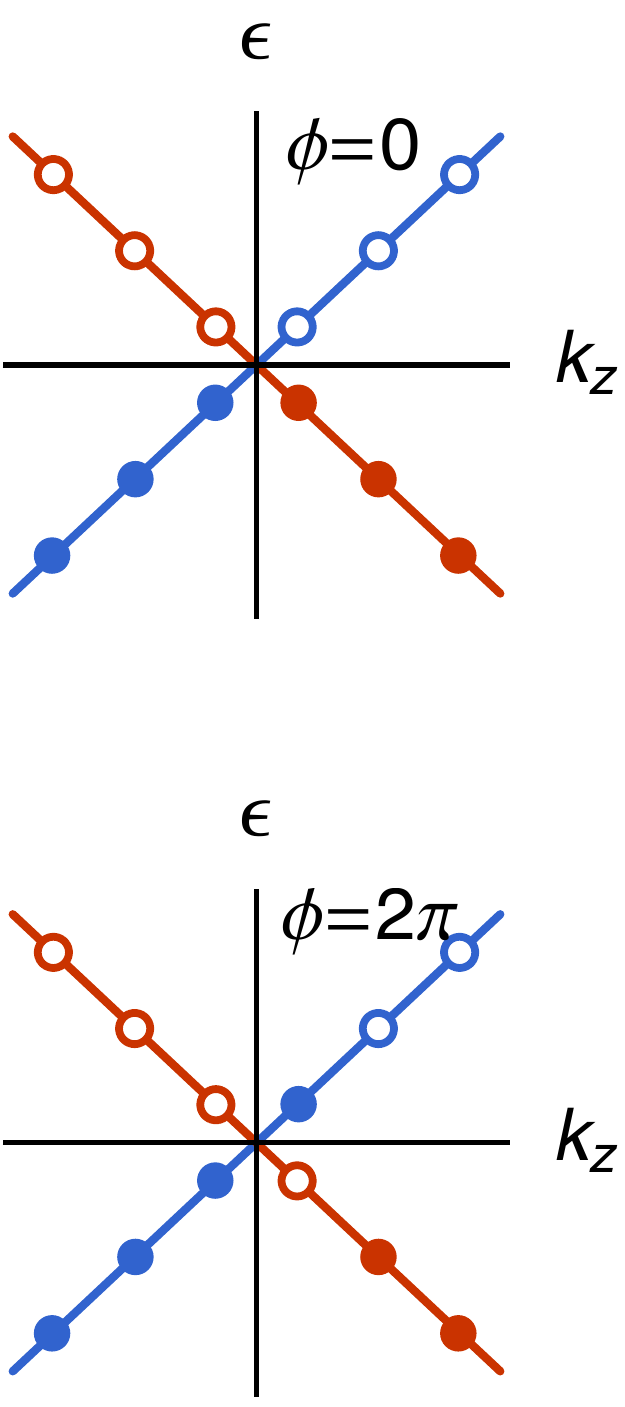}
\includegraphics[width=0.25\textwidth]{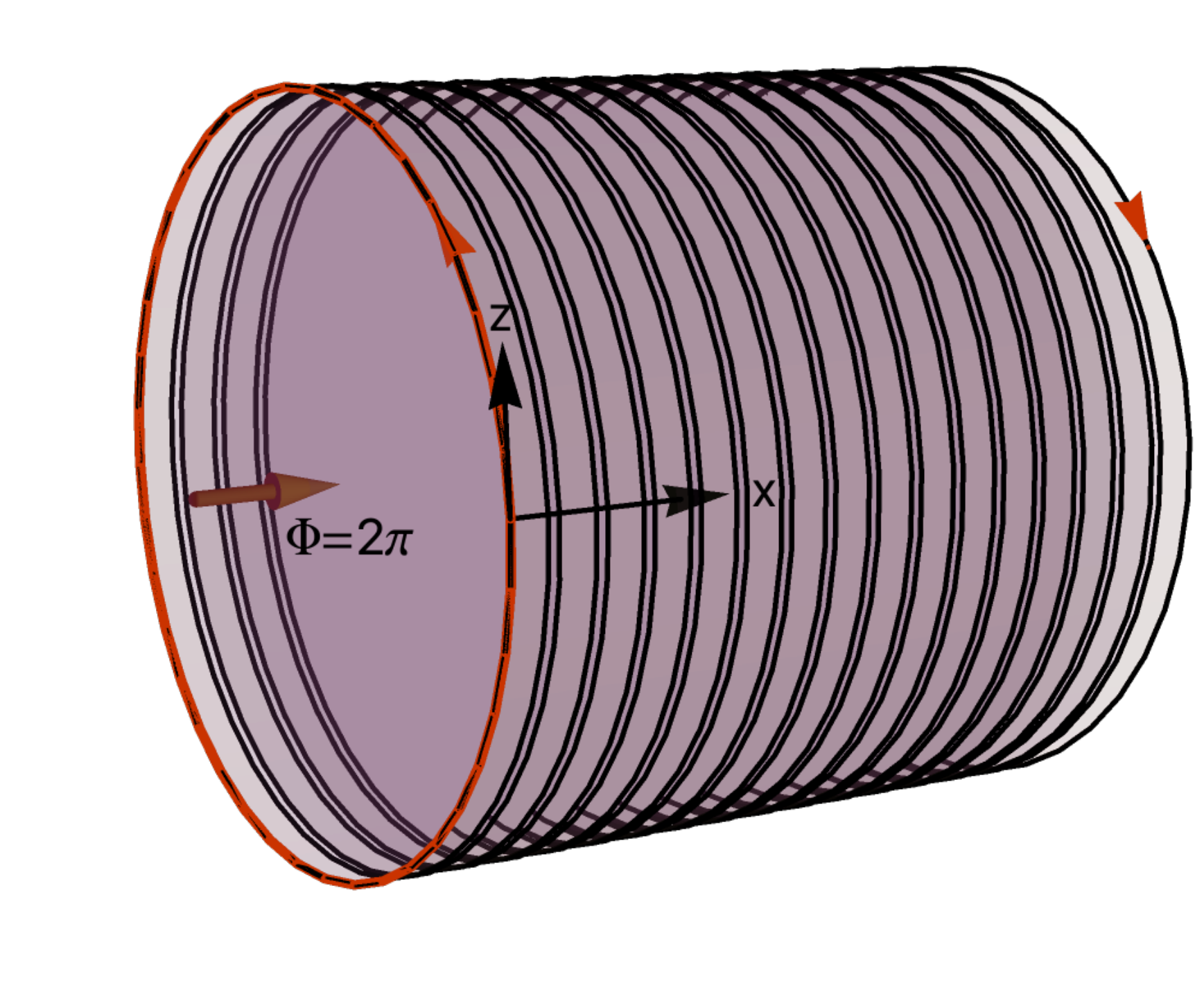}
\caption{A $2$D system with helical edge modes, and U$(1)$ charge conservation along each $z$-row. Periodic boundary conditions are taken along the $z$-direction, and open boundary conditions are taken along the $x$-direction. Since charge is conserved on all $z$-rows, it is possible to insert a subsystem $\Phi = 2\pi$ flux along the leftmost row only. }
\label{anomaly}
\end{figure}

To exhibit this anomaly, we can use a flux threading argument. Since the U(1) charge on the 2D $xz$-plane is conserved on each $z$-row, we can insert a {\it{subsystem}} U(1) flux by inserting a $2\pi$ flux in the cycle spanned by only the leftmost row without affecting the others, as shown in Fig. \ref{anomaly}. As the leftmost row contains two helical modes, such a flux insertion would create a U(1) charge from one chiral mode (which also carries a unit of global $\mathbb{Z}_2$ charge) and creates a U(1) anti-charge from the counter-propagating chiral mode (which is neutral under $\mathbb{Z}_2$). Thus, while the total U(1) charge is conserved, the total $\mathbb{Z}_2$ charge is shifted by a unit under a large gauge transformation of the subsystem U(1) symmetry. This phenomenon signifies a mixed anomaly between the two symmetries, and implies that such a $2$D lattice model in the current setting cannot exist. This also concludes that there is no way to decorate a 2D lattice model with subsystem symmetry on the side surfaces to obtain the same helical hinge states. Note that this is different from the usual physics of the QSH edge state. For a $2$D QSH state with a global U(1) charge symmetry, the flux insertion operator will apply to all degrees of freedom across all rows, so both the left and right edges would respond to such a large gauge transformation. In this case, the combined anomaly from the two edges cancels. However, in the presence of subsystem symmetry, the flux insertion on each row is independent, and the 
large gauge transformation we apply to one edge does not affect the other edge.

\begin{figure}
\centering
\includegraphics[width=0.3\textwidth]{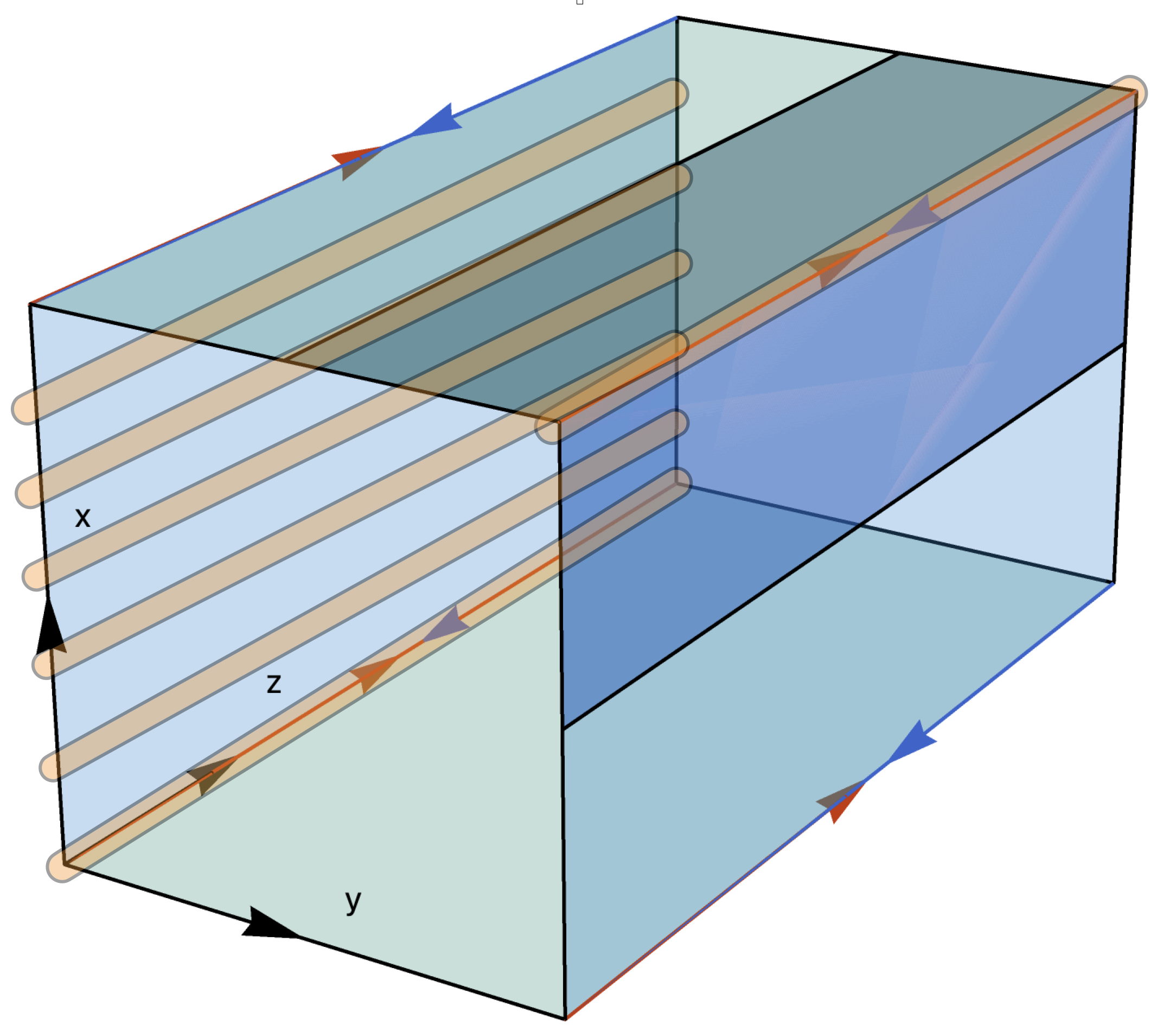}
\caption{The four side surfaces form a tube-shaped cover with four gapless hinges. To demonstrate the anomaly on this side surface, we take out a quadrant of the cover denoted by the darker blue part. This quadrant turns out to be anomalous under certain large gauge transformations by inserting subsystem $ 2\pi $ flux to the rows (yellow lines).}
\label{anomaly2}
\end{figure}
Now let us consider a complementary argument. If we take the whole 2D surface cover of our system it forms a tube-like shape as shown in Fig. \ref{anomaly2}. This cover harbors four helical hinge modes. Additionally, the cover respects the global $\mathbb{Z}_2$ symmetry and the two subsystem U(1) symmetries (the latter act on rows parallel to the $z$-direction on the cover). When we apply a subsystem U(1) flux insertion operator, it will affect two hinges.  For our argument, let us consider an explicit large gauge transformation (LGT) specified by $\theta(r) \rightarrow \theta(r)+\frac{2\pi}{L_z}[\delta(x)\delta(y-N_y)-\sum^{N_x}_{i=1}\delta(x-i)\delta(y)]$ with $\theta$ being the U(1) phase of the compact boson. Such a LGT was chosen so that it involves a global shift of charge from two hinges at $(0,N_y),(N_x,0).$ We see that this LGT does not change the total $\mathbb{Z}_2$ charge on the cover since the hinges change oppositely. More precisely, it inserts a $2\pi$ flux for the helical states at the top-left corner at $(0,N_y)$ and a $(-2\pi)$ flux for the helical states at the bottom-right corner at $(N_x,0)$. Nevertheless, the theory on the cover is still anomalous. To elucidate this, we further cut the cover and extract a quadrant as shown in Fig.~\ref{anomaly2}. The quadrant contains only one hinge and creates new "edges" on the $xz$- and $yz$-surfaces. The new edges might result in additional gapless modes. However, since the side surfaces are fully gapped and short-ranged correlated, edge states on the new edges would not affect the stability of the helical modes on the remaining hinge (which we take to be far away compared to the correlation length). Now if we apply the same subsystem U(1) LGT to the isolated quadrant, the $\mathbb{Z}_2$ charge is not invariant under a large gauge transformation (note that our chosen LGT does not affect the new edges created by the quadrant). This suggests that the 2D theory on the cover is still anomalous since it contains an obstruction in the presence of a boundary.

\subsection{Anomalies in the Entanglement Hamiltonian}
In this subsection, we will demonstrate that the topological nature of the HOSPT can also be probed using the entanglement properties of the wave-functions of the many-body system. In general, the entanglement spectrum of a symmetry protected topological phase does not necessarily resemble the low-energy part of the spectrum at the edge\cite{chandran2014universal}. For example, one can observe phase transitions in the entanglement Hamiltonian that are not reflected in the underlying physics of the ground state wave function of the Hamiltonian. As a result, when we examine the entanglement Hamiltonian of the ground state wave function, we will focus on the whole spectrum rather than only the low-energy states.

To set the stage, we begin with the HOSPT in Eq. \ref{eq:LatticeHamiltonianHOTI}. Since this model is obtained from the coupled wire construction, the ground state wavefunction has (effectively) zero correlation length transverse to the wires. Consider the 3D system with periodic boundary conditions. Let us now choose a subregion $A$ which is open with finite extent in the $x$ and $y$ directions, while is still periodic in the $z$-direction.  The entanglement surface resembles a 2D sheet covering the external side surfaces of region A similar to Fig.~\ref{anomaly2}. From the coupled wire construction, it is straightforward to see that the entanglement Hamiltonian on the side surfaces, i.e., on the $xz$- and $yz$-planes is generically gapped since each row (with fixed coordinate $(x_0,y_0)$) along the $z$-direction contains two sets of decoupled of helical Luttinger liquids that can be coupled without breaking the global $\mathbb{Z}_2$ charge conservation. The four hinges of region $A$ contain additional helical modes where counter-propagating modes have different $\mathbb{Z}_2$ charge assignments. Hence, the hinges remain gapless if the $\mathbb{Z}_2$ global symmetry is preserved. As such, in this fine-tuned limit, the entanglement Hamiltonian looks exactly like the side surface of the 3D HOSPT in Fig.~\ref{anomaly2}. 

While ultimately the gapless nature of the entanglement Hamiltonian could be non-universal, the utility of this calculation is that such a gapless structure indicates the entanglement Hamiltonian contains a symmetry anomaly. This anomaly is exactly analogous to the quantum anomaly of the actual surface, and would hence indicate that such a surface cannot be generated in a lower-dimensional system. The argument for the anomaly is identical to the discussion in Sec. \ref{ssec:HingeAnom}. That is, if we make an additional spatial cut of the entanglement Hamiltonian by keeping only a quadrant of region $A$, then the resulting quadrant contains a mixed anomaly where a large subsystem U(1) gauge transformation on the row containing the hinge will change the $\mathbb{Z}_2$ charge of the system. Thus a gauge transformation of the subsystem U(1) would break the global $\mathbb{Z}_2$ charge conservation, and the resultant entanglement Hamiltonian is anomalous that cannot be realized in a lower-dimensional lattice model with the same symmetry. 

If we move away from the zero-correlation length limit, the helical Luttinger liquids from different wire bundles in the system can interact and couple. While the low-energy spectrum of the entanglement Hamiltonian will change and vary depending on the microscopic couplings, we expect the mixed anomaly to be robust against any wave function (or entanglement Hamiltonian) reconstruction as long as the bulk gap and symmetries are maintained. Subsequently, we expect the mixed anomaly to be a distinguishing feature of the HOSPT entanglement Hamiltonian. 

Finally, let us mention that while the symmetry anomaly persists as long as symmetry is maintained, we cannot ignore the possibility that the entanglement Hamiltonian might have spontaneous symmetry breaking. As the entanglement Hamiltonian is defined in $2$D effectively, the Mermin-Wagner theorem excludes the possibility of subsystem U(1) symmetry breaking\cite{distler2021spontaneously}. However, the global $\mathbb{Z}_2$ symmetry could be broken spontaneously. Such a $\mathbb{Z}_2$ symmetry breaking would then generate a 2-fold degeneracy in the entanglement spectrum.

\section{Higher Order Topological Superconductor with Chiral Hinge States}\label{sec:HOTSC}

In this section, we propose a higher-order topological superconductor (HOTSC) with chiral hinge states that are protected by subsystem $\mathbb{Z}_2$ symmetry. We still adapt the same coupled wire construction formalism as before, and choose interactions such that the model again decomposes into an array of decoupled $1$D bundles. 
\subsection{Fermionic Wire Model}
For this model, we use a unit cell composed of four spinless $1$D Dirac fermions. Again, for simplicity, we assume these Dirac fermions each correspond to excitations near the same lattice momentum. These Dirac fermions can be written in terms of eight complex chiral fermions, $\chi^1_L$, $\chi^2_L$, $\chi^3_R$, $\chi^4_R$, $\chi^5_L$, $\chi^6_L$, $\chi^7_R$, $\chi^8_R$, where, as before, $R$ and $L$ indicate that the mode propagates along the $+z$ and $-z$-directions respectively. Since the net chirality of these modes vanishes, these modes can arise from a microscopic fermionic lattice model.  

The Hamiltonian describing the HOTSC we consider is
\begin{equation}\begin{split}
    &\mathcal{H}_{\text{HOTSC}}=\sum_{\bm{r}} \bm{\chi}^\dagger_{\bm{r}} i\partial_z \tau^{0z0}\bm{\chi}_{\bm{r}} + \mathcal{H}_{\text{int-SC}},\\
    &\mathcal{H}_{\text{int-SC}} = \sum_{\bm{r}} \Big[V_1 \chi^{1 \dagger}_{ L, \bm{r}}\chi^{2\dagger}_{ L,\bm{r}'}\chi^3_{ R,\bm{r}''}\chi^4_{ R,\bm{r}'''}\\
    &\phantom{=======} + V_2 \chi^{5 \dagger}_{L, \bm{r}}\chi^{6\dagger}_{L,\bm{r}'}\chi^7_{R,\bm{r}''}\chi^8_{R,\bm{r}'''}\\
    &\phantom{=======} + V_3 \chi^{1 \dagger}_{L, \bm{r}}\chi^3_{R,\bm{r}''}\chi^{ 5}_{L, \bm{r}}\chi^{7 \dagger}_{R,\bm{r}''}\\
    &\phantom{=======} + V_4 \chi^{1 \dagger}_{L, \bm{r}}\chi^4_{R,\bm{r}'''}\chi^{ 5 \dagger}_{L, \bm{r}}\psi^{8 }_{R,\bm{r}'''}\Big] + h.c., \\
\label{eq:LatticeHamiltonianHOTSC}\end{split}\end{equation}
where  $\bm{\chi} = (\chi^1_L, \chi^2_L,\chi^3_R,\chi^4_R,\chi^5_L,\chi^6_L,\chi^7_R,\chi^8_R)$, the spatial coordinate definitions are the same as those following Eq. \ref{eq:LatticeHamiltonianHOTI}, and we have kept the tensor product notation for the $\tau$ matrices. The $\mathbb{Z}_2$ subsystem symmetries are given by
\begin{equation}\begin{split}
\mathbb{Z}_2^{xz}: \bm{\chi}_{\bm{r}} \rightarrow e^{i \eta_{xz}(\bm{r}\cdot\hat{y})} \bm{\chi}_{\bm{r}},\\
\mathbb{Z}_2^{yz}: \bm{\chi}_{\bm{r}} \rightarrow e^{i \eta_{yz}(\bm{r}\cdot\hat{x})}\bm{\chi}_{\bm{r}},
\label{eq:SubSystemDef}\end{split}\end{equation}
where $\eta_{xz}$ and $\eta_{yz}$ are functions of $\bm{r}\cdot\hat{y} = n_y$ and $\bm{r}\cdot\hat{x} = n_x$ respectively, and are $\{0,\pi\}$ valued. 

Similar to before, the eight fermions $\chi^{ 1}_{L, \bm{r}}$, $\chi^{2}_{L,\bm{r}'}$, $\chi^3_{ R,\bm{r}''}$, $\chi^4_{R,\bm{r}'''}$, $\chi^{ 5}_{L, \bm{r}}$, $\chi^{6}_{L,\bm{r}'}$, $\chi^7_{R,\bm{r}''}$, and $\chi^8_{R,\bm{r}'''}$ only couple to one another in a bundle for fixed $\bm{r}$. To show that the interactions in Eq. \ref{eq:LatticeHamiltonianHOTSC} gap out the bulk fermions, we will once again use bosonization. Here, the complex fermions modes correspond to the vertex operators $\chi^j_{R/L} \sim e^{\mp i \varphi^j_{R/L}}$, where the $\mp$ are correlated to the $R/L$ subscript, and $j=1,\ldots 8.$
In terms of these bosonic fields, the interactions in Eq. \ref{eq:LatticeHamiltonianHOTSC} become
\begin{equation}\begin{split}
\mathcal{H}_{\text{int-SC}} = &-g_1 \cos(\varphi^1_{L,\bm{r}} + \varphi^2_{L,\bm{r}'} + \varphi^3_{R,\bm{r}''} + \varphi^4_{R,\bm{r}'''})  \\ & -g_2 \cos(\varphi^5_{L,\bm{r}} + \varphi^6_{L,\bm{r}'} + \varphi^7_{R,\bm{r}''} + \varphi^8_{R,\bm{r}'''}) \\ & -g_3 \cos(\varphi^1_{L,\bm{r}} + \varphi^3_{R,\bm{r}''} -\varphi^5_{L,\bm{r}}-\varphi^7_{R,\bm{r}''})  \\ & -g_4 \cos(\varphi^1_{L,\bm{r}} + \varphi^4_{R,\bm{r}'''} + \varphi^5_{L,\bm{r}} + \varphi^8_{R,\bm{r}'''}).
\end{split}\end{equation}
These terms all commute with each other, and hence each wire bundle in the bulk is gapped at strong coupling. 

Having seen that the bulk is gapped we can consider surface boundaries normal to the $x$ or $y$-directions. On such boundaries there exist gapless fermionic modes with vanishing chirality which can subsequently be gapped by turning on a surface coupling. Finally, on the hinges, e.g., the hinge at the intersection between the surfaces with $+\hat{x}$ and $+\hat{y}$ normal vectors, the fermions $\chi^{ 1}_{L}$, $\chi^3_{ R}$, $\chi^4_{R}$, $\chi^{ 5}_{L}$, $\chi^7_{R}$, and $\chi^8_{R}$ are gapless. Hence, there are 2 stable chiral complex fermion modes (equiv. 4 real Majorana modes) at each hinge and this system represents a HOTSC. 

The chiral hinges modes we described here are subject to surface modifications and reconstruction.  Importantly, the number of Majorana hinge modes can be changed by adding a $2$D topological superconductor with $\mathbb{Z}_2$ subsystem symmetry to the surface. In Appendix \ref{App:2DTSC}, we show that the chiral Majorana edge modes of a $2$D topological superconductor with $\mathbb{Z}_2$ subsystem symmetry come in multiples of $8$. Hence, the number of chiral Majorana hinge modes is therefore only defined modulo $8$ for a $3$D $\mathbb{Z}_2$ subsystem symmetric insulator. Based on this, a single copy of the HOTSC--with $4$ Majorana hinge modes--is non-trivial, while two copies of the HOTSC--with $8$ Majorana hinge modes--are trivial. The HOTSC therefore has a $\mathbb{Z}_2$ classification. As a corollary, we see that the chirality of the hinge modes can be flipped by surface modifications. 

\subsection{Hinge Anomaly}
The same argument used in Sec. \ref{ssec:HingeAnom} to show that the helical state of the HOSPT is anomalous can be adapted to show that the chiral hinge state of the HOTSC is also anomalous. Here, we take a gapped side surface on the $xz$-plane with two sets of $c=2$ chiral edges states along $z$.
Since the $\mathbb{Z}_2$ fermion parity is conserved on each row parallel to $z$ at fixed $(x,y)$, the two hinges have independent fermion parity charges denoted as $\mathbb{Z}^L_2$ and $\mathbb{Z}^R_2$. As a whole, the counter propagating hinge modes from the left and right part of the $xz$-plane side surface resemble the edge physics of four copies of a 2D $p \pm ip$ SC  where the left moving modes and right moving modes have different $\mathbb{Z}_2$ charges\cite{you2015bridging}\footnote{If we squash the side surface in the $x$-direction, we end up with an edge resembling four copies of a combined $p+ip$ and $p-ip$ superconductor.}. Such a surface theory contains a global $\mathbb{Z}_2$ anomaly, so the corresponding side surface cannot be realized in a pure 2D subsystem symmetric system. In the Appendix.~\ref{App:2DTSC}, we provide a detailed argument to demonstrate that 2D superconductors with subsystem $\mathbb{Z}_2$ symmetry must carry a minimum of eight chiral Majorana edge modes, i.e., a central charge of $c=4$, which is twice the amount on the hinges of this 3D HOTSC.

\subsection{Comparison to U$(1)$ Subsystem Symmetry}\label{ssec:Z2andU1}
When considering the HOSPT in Sec. \ref{ssec:NLSM}, we noted that U(1) subsystem symmetries acted on the set of fermions $\tilde{\psi}_\uparrow$ in the same manner as global U$_c(1)\times$U$_s(1)$ symmetry acting on the edge of a 2D QSH SPT. The classification of such 2D QSH SPTs is $\mathbb{Z}$, and because of this, it was necessary to include a second quartet of fermions with the \textit{opposite} topological index, i.e., the $\tilde{\psi}_\downarrow$ fermions, in order to fully gap the bulk. Because the $\tilde{\psi}_\uparrow$ and $\tilde{\psi}_\downarrow$ fermions have opposite $\mathbb{Z}$ topological indices, this provides a simple way to argue that the hinge modes of the HOSPT we constructed are helical. 

Interestingly, for the HOTSC case, we find chral modes on the hinges instead of helical modes. To illustrate how this occurs, we can adapt the previous argument to the case of the HOTSC where we have an analogous set of fermions $\tilde{\chi}_{a,\bm{r}} = (\chi^{ 1}_{L, \bm{r}}, \chi^{2}_{L,\bm{r}'}, \chi^3_{ R,\bm{r}''}, \chi^4_{R,\bm{r}'''})$. Through a mechanism similar as in Eq. \ref{eq:hinge} and Eq. \ref{eq:NLSM1} in Sec. \ref{ssec:NLSM}, this set of fermion modes can be reduced to an O(4) NLSM with a WZW term. Here, the $\mathbb{Z}_2$ subsystem symmetry actions on this NLSM precisely map to the edge of a 2D bosonic SPT with $\mathbb{Z}_2 \times \mathbb{Z}_2$ symmetry. Crucially, such an SPT has a $\mathbb{Z}_2$ classification, hence the $\tilde{\chi}_{a,\bm{r}}$ fermions can be gapped if we just add an identical set of fermions with the \textit{same} $\mathbb{Z}_2$ classification. Such a set is given by  $\tilde{\chi}_{b,\bm{r}} = (\chi^{ 5}_{L, \bm{r}}, \chi^{6}_{L,\bm{r}'}, \chi^7_{R,\bm{r}''}, \chi^8_{R,\bm{r}'''})$.  From this, we can conclude that the bulk of the HOTSC can be consistently gapped. (An alternative way to see the gapped bulk from a topological defect perspective is provided in Appendix \ref{defect}.) Furthermore, based on how the bundles $\tilde{\chi}_{a,\bm{r}}$ and $\tilde{\chi}_{b,\bm{r}}$ are embedded in the $3$D wire construction, we can also conclude that the HOTSC will have $4$ chiral Majorana modes on each hinge.

We can also generalize this idea to systems with discrete $\mathbb{Z}_N$ subsystem symmetry. Let us start by considering a system with $\mathbb{Z}_N$ subsystem symmetry composed of 4$N'$ complex fermion wires per unit cell: 2$N'$ right moving and 2$N'$ left moving. These wires can be combined into $N'$ bundles of $4$ wires--2 right moving and 2 left moving. Each bundle with the $\mathbb{Z}_N$ subsystem symmetry carries the same anomaly as the edge of a bosonic SPT with $\mathbb{Z}_N\times \mathbb{Z}_N$ symmetry. The classification of such SPTs is $\mathbb{Z}_N$, and so the $N'$ bundles can be symmetrically gapped only when $N'$ is a multiple of $N$. For $N' = N$, one can explicitly write down the gapping term in the bosonization language, and the resulting gapped system will be non-trivial higher order phase hosting $N$ chiral complex fermionic modes on each hinge.

\begin{figure}[h]
    \centering
    \includegraphics[width=0.2\textwidth]{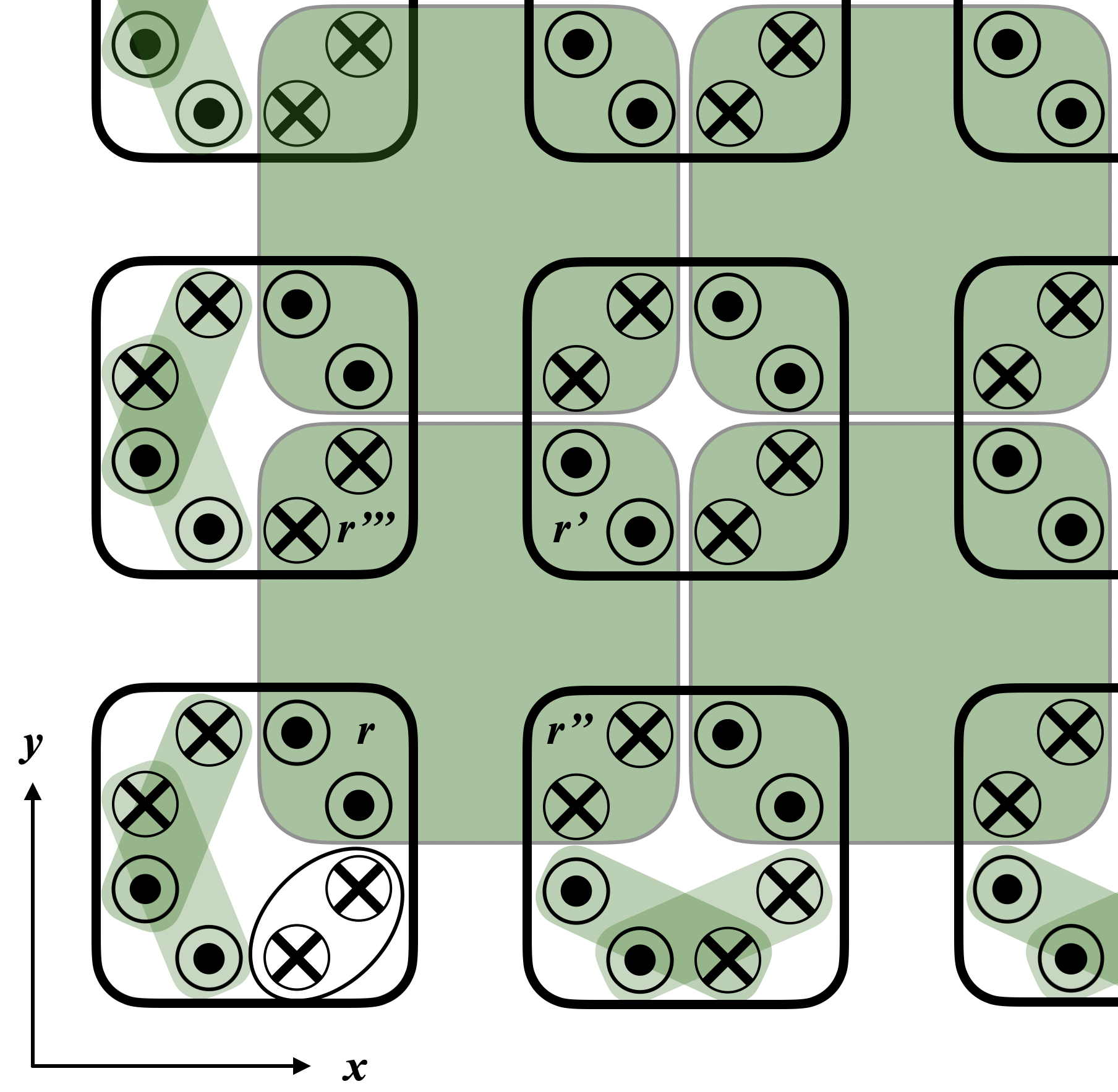}
    \includegraphics[trim=-1in 0 0 0, clip, width=0.25\textwidth]{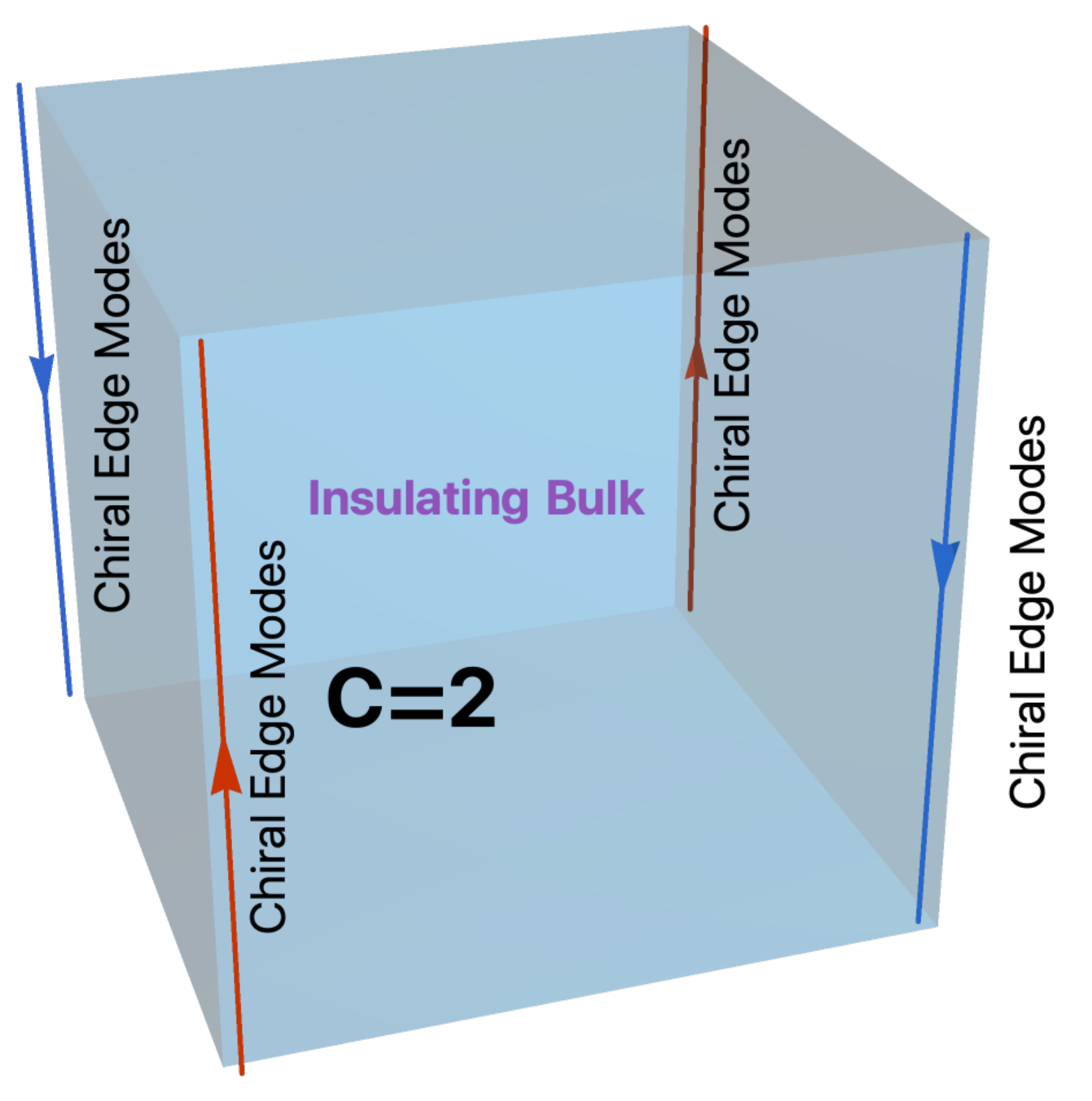}
    \caption{(Left) Wire construction for the higher order topological superconductor with chiral hinge modes. (Right) The schematic of chiral modes on the hinges of the system. For this construction, the chiral central charge of the hinge mode is 2, i.e., it consists of four chiral Majorana modes.}
    \label{chiral}
\end{figure}

\section{Fractionalized Higher Order Topological Insulator}\label{sec:FHOTI}
In this section we propose a fractional higher-order topological insulator (FHOTI) with helical hinge modes that are protected by U$(1)$ subsystem symmetry, and a global $\mathbb{Z}_2$ symmetry. These helical hinge modes have fractional charge $1/m = 1/(2n+1)$, $n\in \mathbb{Z}$ under the U$(1)$ subsystem symmetries. As might be expected from the fractionalized hinge modes, the FHOTI also has a topological ground state degeneracy. The ground state degeneracy scales subextensively with system size, similar to what is seen in fractonic phases of matter\cite{vijay2016fracton}. 

\subsection{Fermionic Wire Model}
Similar to before, we construct the FHOTI using a coupled wire formalism. For the FHOTI construction, the low-energy degrees of freedom of the wires consist of eight spin up and eight spin down $1$D fermions, which we label as $\psi^i_{R/L,\sigma}$ for $i = 1...4$. The $\psi^1_{R,\uparrow}$, $\psi^2_{R,\uparrow}$, $\psi^3_{R,\downarrow}$, and $\psi^4_{R,\downarrow}$ fermions correspond to excitations near lattice momentum $k_R + b/2$. Similarly, $\psi^1_{L,\uparrow}$, $\psi^2_{L,\uparrow}$, $\psi^3_{L,\downarrow}$, and $\psi^4_{L,\downarrow}$ are excitations near momentum $k_L + b/2$, $\psi^1_{R,\downarrow}$, $\psi^2_{R,\downarrow}$, $\psi^3_{R,\uparrow}$, and $\psi^4_{R,\uparrow}$ are excitations near momentum $k_R-b/2$, and $\psi^1_{L,\downarrow}$, $\psi^2_{L,\downarrow}$, $\psi^3_{L,\uparrow}$, and $\psi^4_{L,\uparrow}$ are excitations near momentum $k_L-b/2$. The constant $b$ is defined to satisfy $ b = m(k_R-k_L)$ with $m \in \mathbb{Z}$. 

The Hamiltonian for the FHOTI is given by
\begin{equation}\begin{split}
&\mathcal{H}_{\text{FHOTI}} = \sum_{\bm{r}} \bm{\Psi}^\dagger_{\bm{r}} i \partial_z  \tau^{00z0} \bm{\Psi}_{\bm{r}}  + \mathcal{H}_{\text{intra}} + \mathcal{H}_{\text{inter}},\\
&\mathcal{H}_{\text{intra}} = \sum_{\bm{r}} J_1 \mathcal{O}^{1\dagger}_{R,\uparrow,\bm{r}} \mathcal{O}^{3}_{L,\uparrow,\bm{r}} +  J_2 \mathcal{O}^{2\dagger}_{R,\uparrow,\bm{r}} \mathcal{O}^{4}_{L,\uparrow,\bm{r}}\\
&\phantom{======}+J_3 \mathcal{O}^{1\dagger}_{L,\downarrow,\bm{r}} \mathcal{O}^{4}_{R,\downarrow,\bm{r}} +  J_4 \mathcal{O}^{2\dagger}_{L,\downarrow,\bm{r}} \mathcal{O}^{3}_{R,\downarrow,\bm{r}}\\
&\mathcal{H}_{\text{inter}} = \sum_{\bm{r}} V_1 \mathcal{O}^{1\dagger}_{L,\uparrow,\bm{r}} \mathcal{O}^{2\dagger}_{L,\uparrow,\bm{r}'} \mathcal{O}^{3}_{R,\uparrow,\bm{r}''} \mathcal{O}^{4}_{R,\uparrow,\bm{r}'''}\\
&\phantom{======}+V_2 \mathcal{O}^{1\dagger}_{R,\downarrow,\bm{r}} \mathcal{O}^{2\dagger}_{R,\downarrow,\bm{r}'} \mathcal{O}^{3}_{L,\downarrow,\bm{r}''} \mathcal{O}^{4}_{L,\downarrow,\bm{r}'''}\\
&\phantom{======}+V_3 \mathcal{O}^{1\dagger}_{L,\uparrow,\bm{r}} \mathcal{O}^{3}_{R,\uparrow,\bm{r}''}\mathcal{O}^{1}_{R,\uparrow,\bm{r}} \mathcal{O}^{3\dagger}_{L,\uparrow,\bm{r}''}\\
&\phantom{======}+V_4 \mathcal{O}^{1\dagger}_{L,\uparrow,\bm{r}} \mathcal{O}^{4\dagger}_{R,\uparrow,\bm{r}'''}\mathcal{O}^{1}_{R,\downarrow,\bm{r}} \mathcal{O}^{4\dagger}_{L,\downarrow,\bm{r}'''},\\
&\mathcal{O}^{i}_{R/L,\sigma,\bm{r}} = (\psi^{i\dagger}_{L/R,\sigma,\bm{r}}\psi^{i}_{R/L,\sigma,\bm{r}})^n \psi^{i}_{R/L,\sigma,\bm{r}}
\label{eq:FHOTILatticeHam}\end{split}\end{equation}
where $\bm{\Psi} = (\psi^1_{R,\uparrow}...\psi^4_{R,\uparrow}$, $\psi^1_{L,\uparrow}...\psi^4_{L,\uparrow}$, $\psi^1_{R,\downarrow}...\psi^4_{R,\downarrow}$, $\psi^1_{L,\downarrow}...\psi^4_{L,\downarrow} )$. Provided that $b = m(k_R-k_L)$ all interactions in Eq. \ref{eq:FHOTILatticeHam} carry vanishing momentum. Eq. \ref{eq:FHOTILatticeHam} has two U$(1)$ subsystem symmetries, which are defined analogously to those in Eq. \ref{eq:SubSystemDef}. There is also a global $\mathbb{Z}_2$ symmetry that sends $\bm{\Psi} \rightarrow \tau^{000z} \bm{\Psi}$, and, similarly, $\mathcal{O}^{i}_{R/L,\uparrow} \rightarrow \mathcal{O}^{i}_{R/L,\uparrow}$ and $\mathcal{O}^{i}_{R/L,\downarrow} \rightarrow -\mathcal{O}^{i}_{R/L,\downarrow}$.

To analyze the interacting Hamiltonian in Eq. \ref{eq:FHOTILatticeHam} we shall use bosonization, identifying $\psi^i_{R/L,\sigma} \sim \exp( \mp i \phi^i_{R/L,\sigma})$, where the $\mp$ correlate with the $R/L$ subscript. To simplify the interactions, we define the following bosons:
\begin{equation}\begin{split}
&\tilde{\phi}^i_{R\sigma} = \frac{n+1}{m}\phi^i_{R \sigma} + \frac{n}{m}\phi^i_{L \sigma},\\
&\tilde{\phi}^i_{L\sigma} = \frac{n+1}{m}\phi^i_{L \sigma} + \frac{n}{m}\phi^i_{R \sigma}. 
\label{eq:FHOTfracBosons}\end{split}\end{equation}
These bosons satisfy the commutation relationships 
\begin{equation}\begin{split}
[\tilde{\phi}^i_{R/L \sigma}(z), \tilde{\phi}^j_{R/L \sigma'}(z')] = \pm \pi \frac{1}{m} \delta_{\sigma,\sigma'}\delta_{ij}\text{sgn}(z-z').
\label{eq:fracCom}\end{split}\end{equation}
These are exactly the commutation relationships of the surface modes of a Laughlin quantum Hall state at filling $1/m$. The U(1) charge operator for the bosons is $\rho = \frac{1}{2\pi} \sum_{i,\sigma} \partial_z [\tilde\phi^i_{R,\sigma} +  \tilde\phi^i_{L,\sigma}]$, and the vertex operators $\exp(i \tilde{\phi}^i_{R/L \sigma})$ carry charge $\pm 1/m$. 
In terms of these new bosonic fields $\mathcal{O}^{i}_{R/L,\sigma,\bm{r}} \sim \exp(\mp i m \tilde{\phi}^i_{R/L \sigma})$. 

With this in mind, let us first consider the interactions in $\mathcal{H}_{\text{intra}}$:
\begin{equation}\begin{split}
\mathcal{H}_{\text{intra}} = \sum_{\bm{r}} &\lambda_1 \cos(m[\tilde{\phi}^1_{R,\uparrow,\bm{r}}+ \tilde{\phi}^3_{L,\uparrow,\bm{r}}])\\
&+\lambda_2 \cos(m[\tilde{\phi}^2_{R,\uparrow,\bm{r}}+ \tilde{\phi}^4_{L,\uparrow,\bm{r}}])\\
&+\lambda_3 \cos(m[\tilde{\phi}^1_{L,\downarrow,\bm{r}}+ \tilde{\phi}^4_{R,\downarrow,\bm{r}}])\\
&+\lambda_4 \cos(m[\tilde{\phi}^2_{L,\downarrow,\bm{r}}+ \tilde{\phi}^3_{R,\downarrow,\bm{r}}]).\label{eq:FHOTIintraBoson}
\end{split}\end{equation}
When the $\lambda_i$ couplings are large, the only gapless fields are $\tilde{\phi}^{1/2}_{L \uparrow}$, $\tilde{\phi}^{1/2}_{R \downarrow}$ ,$\tilde{\phi}^{3/4}_{R \uparrow}$, and $\tilde{\phi}^{3/4}_{L \downarrow}$. The intrawire interactions therefore turn each unit cell into 4 sets of helical modes, each with charge $1/m$. A single such unit cell can be described in the K-matrix formalism, using the $8\times 8$ matrix $K = m\tau^{0zz},$ and the $8$ component charge vector $t = (1,...,1)$\cite{wen1992classification}.  

Let us now consider the interwire interactions, $\mathcal{H}_{\text{inter}}$. In terms of the bosonic fields from Eq. \ref{eq:FHOTfracBosons} the interactions in $\mathcal{H}_{\text{inter}}$ are
\begin{equation}\begin{split}
\mathcal{H}_{\text{inter}} & =  g_1 \cos(m[\tilde\phi^1_{\uparrow L,\bm{r}} + \tilde\phi^2_{\uparrow L,\bm{r}'} + \tilde\phi^3_{\uparrow R,\bm{r}''} + \tilde\phi^4_{\uparrow R,\bm{r}'''}])  \\ & +g_2 \cos(m[\tilde\phi^1_{\downarrow R,\bm{r}} + \tilde\phi^2_{\downarrow R,\bm{r}'} + \tilde\phi^3_{\downarrow L,\bm{r}''} + \tilde\phi^4_{\downarrow L,\bm{r}'''}]) \\ & +g_3 \cos(m[\tilde\phi^1_{\uparrow L,\bm{r}} + \tilde\phi^3_{\uparrow R,\bm{r}''} -\tilde\phi^1_{\downarrow R,\bm{r}}-\tilde\phi^3_{\downarrow L,\bm{r}''}])  \\ & +g_4 \cos(m[\tilde\phi^1_{\uparrow L,\bm{r}} + \tilde\phi^4_{\uparrow R,\bm{r}'''}- \tilde\phi^1_{\downarrow R,\bm{r}}- \tilde\phi^4_{\downarrow L,\bm{r}'''}]).
\label{eq:FHOTIinterBoson}\end{split}\end{equation}
When the $g_i$ couplings are large, the bulk bosonic modes are gapped out, and there are gapless hinge degrees of freedom. As noted before, these helical modes have charge $1/m$. In general, these interactions can be made relevant by tuning various symmetry preserving scattering terms in Eq. \ref{eq:FHOTILatticeHam}. We have therefore constructed a fully gapped HOTI with fractionalized symmetry protected hinge modes. It should be noted that the FHOTI reduces to the HOTI of Sec. \ref{sec:HOTI} when $m = 1$ ($n = 0$). Also, this construction can be generalized to produce FHOTIs where the helical hinge modes have other rational charges.

\subsection{Boundary Anomaly}
Similar to the HOSPT of Sec. \ref{sec:HOTI}, the FHOTI presented here has a mixed anomaly between the U$(1)$ subsystem symmetries, and the global $\mathbb{Z}_2$ symmetry. For the FHOTI, inserting a $2\pi$ flux changes the $\mathbb{Z}_2$ charge localized at a hinge by $1/m$. To show this, we use the following effective $1$D Lagrangian to describe the fractionalized hinge mode of the FHOTI:
\begin{equation}\begin{split}
\mathcal{L}_{\text{FHOTI-Hinge}} = &\frac{m}{4\pi}\partial_t \tilde{\bm{\phi}}^T \tau^z \partial_z \tilde{\bm{\phi}} - \frac{1}{4\pi} \partial_z \tilde{\bm{\phi}}^T V \partial_z \tilde{\bm{\phi}}\\ &- \frac{1}{2\pi} \epsilon_{\mu\nu} \bm{t}\cdot \tilde{\bm{\phi}} \partial_\mu A_\nu,
\label{eq:HingeEff}\end{split}\end{equation}
where $\tilde{\bm{\phi}} = (\tilde{\phi}_{R,\uparrow},\tilde{\phi}_{L,\downarrow})$ is a two component boson that encodes the fractionalized hinge mode, $\bm{t} = (1,1)$ is the charge vector, and $V$ is a $2\times 2$ velocity matrix. We have also gauged the U$(1)$ subsystem symmetry and coupled the hinge to the U$(1)$ gauge field $A_{\mu}$ ($\mu = t,z$).  The $\mathbb{Z}_2$ global symmetry acts via $\tilde{\bm{\phi}} \rightarrow \tilde{\bm{\phi}} + \pi \bm{\alpha}$, where $\bm{\alpha} = (0,1/m)$ and the $\mathbb{Z}_2$ charge and current densities are $j^t_{\mathbb{Z}_2} = \frac{m}{2\pi}\bm{\alpha}^T \tau^z \partial_z \tilde{\bm{\phi}}$ and $j^z_{\mathbb{Z}_2} = \frac{1}{2\pi}\bm{\alpha}^T V \partial_z \tilde{\bm{\phi}}$. For $m=1$ this theory describes the non-fractionalized helical hinge mode of the HOSPT.

Using the equations of motion from Eq. \ref{eq:HingeEff}, we find the following anomalous conservation equation for the $\mathbb{Z}_2$ charge and current,
\begin{equation}\begin{split}
\partial_\mu j^\mu_{\mathbb{Z}_2}  = \frac{1}{2\pi} \bm{\alpha}\cdot \bm{t} \epsilon_{\mu\nu} \partial_\mu A_\nu.
\label{eq:Z2FracAnom}\end{split}\end{equation}
Hence, if we insert a $2\pi$ U$(1)$ flux, the $\mathbb{Z}_2$ charge increases by $\bm{\alpha}\cdot \bm{t} = 1/m$. As before, for $m = 1$, we find that inserting $2\pi$ flux increases the $\mathbb{Z}_2$ charge by $1$. For $m>1$, the change in $\mathbb{Z}_2$ charge is fractional, which reflects the fractionalization of the microscopic fermions. 

Clearly, Eq. \ref{eq:Z2FracAnom} indicates that the hinge mode of the FHOTI cannot occur in a symmetric $1$D system. Additionally, using similar augments to those given in Sec. \ref{ssec:HingeAnom}, the FHOTI hinge mode can also not be realized as the edge mode of a subsystem symmetric $2$D system. In short, let us consider a such a subsystem symmetric $2$D insulator with an edge mode with the same anomaly as in Eq. \ref{eq:Z2FracAnom}. Due to subsystem symmetry, we can consider threading a $2\pi$ U$(1)$ flux \text{only} along the edge of this system. Based on Eq. \ref{eq:Z2FracAnom}, the $\mathbb{Z}_2$ charge of the edge would increase during this process. Since the flux is only threaded at the edge of the system, this process would necessarily increase the $\mathbb{Z}_2$ charge of the full $2$D system as well, indicating that it is anomalous. We therefore conclude that the FHOTI hinge modes are anomalous and cannot be realized at the edge of a purely 2D system with the same symmetries.

\subsection{Ground State Degeneracy}
As one would expect, the FHOTI has topological ground state degeneracy when defined on a lattice with periodic boundary conditions. Specifically, for a $L_x \times L_y \times L_z$ lattice, there are $m^{2L_x + 2L_y-2}$ ground states. This linear scaling of ground state degeneracy is similar to what is seen in fractonic phases of matter\cite{nandkishore2019fractons}. The ground state degeneracy is derived in Appendix \ref{App:GSD} using methodology similar to that of Ref. \onlinecite{meng2020coupled}.

It is worth remarking that each of the 8 cosine terms in Eq. \ref{eq:FHOTIintraBoson} and \ref{eq:FHOTIinterBoson} has $m$ possible minima, which would naively lead to $m^{8LxLy}$ degenerate ground states. However, we also need to take into account that the bosons $\phi^i_{R/L \sigma}$ are compact ($ \phi^i_{R/L \sigma} \equiv \phi^i_{R/L \sigma}+ 2\pi$). So the bosons defined in Eq. \ref{eq:FHOTfracBosons} satisfy $(\tilde{\phi}^i_{R \sigma}, \tilde{\phi}^i_{R \sigma}) \equiv (\tilde{\phi}^i_{R \sigma} + \pi/m, \tilde{\phi}^i_{R \sigma}-\pi/m)$ for each spin and flavor index $i$. Taking this into account, the number of degenerate ground states is reduced to $m^{2L_x + 2L_y-2}$ as shown in Appendix \ref{App:GSD}.

\section{Conclusion and Outlook}
In this work, we presented and analyzed three different microscopic models of subsystem symmetric HOSPTs with gapless hinge modes. We have showed these systems display a number of unique properties, primarily arising from subsystem symmetry. First, the subsystem symmetric HOSPTs are necessarily interacting, and cannot be realized in non-interacting systems. Second, the hinge modes of these models are protected by a combination of subsystem and global symmetries, and are stable without spatial symmetries. To our knowledge, the models we constructed here represent the first microscopic models of chiral HOSPTs that are stable in the absence of any spatial symmetries. Third, the $1$D modes that are localized at the hinges of the subsystem symmetric HOSPTs are anomalous and hence cannot appear in any purely $1$D system, or as the edges of any purely $2$D system. In particular, it is not possible to realize these $1$D modes at the edges of a subsystem symmetric $2$D system. 

This work also raises a number of interesting questions. 1) Is it possible to construct a model where the hinge modes have central charge $c= 1/2$ (i.e., a single chiral Majorana fermion) or $c= 1$ (i.e., a single chiral complex fermion)?
In this work, we have presented models with helical hinge modes (central charge $c = 0$), a model with 4 chiral Majorana hinge modes ($c = 2$), and it remains to be seen if there is any obstruction to realizing HOSPTs with lower central charges. 2) Are there any HOSPTs which are stabilized \textit{only} by U(1) subsystem symmetries? The examples presented here require either a global symmetry, or a $\mathbb{Z}_2$ subsystem symmetry. 3) What subsystem symmetric HOSPTs can be constructed from microscopic bosons? All the models in this paper are constructed out of microscopic fermions (although we used bosonization to describe the low-energy physics in terms of bosons). It is therefore natural to ask what subsystem symmetric bosonic HOSPTs are possible, and if these bosonic HOSPTs are fundamentally different from the fermionic ones.

It is also worth considering possible topological field theory descriptions of these systems. These would likely be related to the dipolar Chern-Simons theory presented in Ref. \cite{you2021multipolar}. However, the dipolar Chern-Simons theory describes a system with U$(1)$ subsystem symmetry,  $c= 1$ hinge modes, and additional gapless surface modes. The models in this work all have gapped boundaries, and either helical hinge modes or $\mathbb{Z}_2$ subsystem symmetry. It seems likely that a multicomponent version of the dipolar Chern-Simons theory could describe the helical HOTI, and a Higgsed version of the dipolar Chern-Simons theory could describe the HOTSC, but we leave further discussion to future work.

\section{Acknowledgements}

TLH and JMM thank ARO MURI W911NF2020166 for support.
 JMM is also supported by the National Science Foundation Graduate Research Fellowship Program under Grant No. DGE - 1746047. YY is supported by H2020-MSCA-IF-2020 and the Gordon Betty Moore foundation. ZB is supported by the startup funding from Penn State University.

\bibliography{SSHO.bib}
\bibliographystyle{apsrev4-1}

 \appendix
 
\section{Defect Proliferation}
\label{defect}
In this appendix, we provide an alternative perspective of a $\mathbb{Z}_2$ subsystem symmetric topological superconductor in terms of topological defects. Essentially, we want to argue that the set of modes $\tilde{\chi}_{a,\bm{r}}$ and $\tilde{\chi}_{b,\bm{r}}$ defined in Sec. \ref{ssec:Z2andU1} are gappable by turning on symmetric interactions. Let us decompose the two quartets of complex fermions from $\tilde{\chi}_{a,\bm{r}}$ and $\tilde{\chi}_{b,\bm{r}}$ into 16 Majorana fermions, which we label as $\gamma_{\bm{r}}= (\gamma_{1,\bm{r}} ... \gamma_{16,\bm{r}})$. For a given $\bm{r}$, the Hamiltonian for $\gamma$ can be written as (leaving the fixed $\bm{r}$ implicit)
\begin{equation}
    H_{m}=\gamma^T(i\partial_z\tau^{z000})\gamma.
\end{equation}
Linear combinations of the $\mathbb{Z}_2$ subsystem symmetry on this set of wires give us 3 independent $\mathbb{Z}_2$ symmetries whose actions are as the following,
\begin{equation}\begin{split}
&\mathbb{Z}_2^{xz-\gamma}: \gamma\rightarrow -\tau^{zz00}\gamma,\\
&\mathbb{Z}_2^{yz-\gamma}: \gamma\rightarrow -\tau^{0z00}\gamma,\\
&\mathbb{Z}_2^{\text{total}-\gamma}: \gamma\rightarrow -\gamma.\\
\end{split}\end{equation}

We will first consider gapping out the system by a $\mathbb{Z}_2^{xz-\gamma}$ symmetry-breaking mass term $m\gamma^T\tau^{y000}\gamma$. Then we will try to restore the symmetry phase by considering dynamical fluctuations of this mass term. This transition to a disordered phase can be viewed as a condensation of topological defects of the $\mathbb{Z}_2$ order parameter, namely the domain walls. To ensure a gapped symmetric phase after the condensation, the topological defect must not carry any nontrivial zero modes or quantum numbers. If such a trivial defect exists, then it is possible to achieve the symmetric gapped bulk state that is desired. 

Let us examine the domain wall of this $\mathbb{Z}_2^{xz-\gamma}$ breaking mass term. It turns out that such a domain wall carries 8 Majorana zero modes. We can label them by $\tilde{\gamma}_l$, $l=1,2,...,8$. The $\mathbb{Z}_2^{yz-\gamma}$ symmetry acts on these modes as $\mathbb{Z}_2^{yz-\tilde{\gamma}}:\tilde{\gamma}\rightarrow \tau^{z00}\tilde{\gamma}$. Our task is to find a symmetric four fermion interaction (two-fermion interactions will violate the subsystem symmetry since our fermions are spread across multiple unit cells) that can gap out the domain wall zero modes and leave a non-degenerate ground state. Let us compose the Majorana zero modes into complex fermions: $f_i=\tilde{\gamma}_i-i\tilde{\gamma}_{i+4}$. In terms of the $f$ fermions, the $\mathbb{Z}_2^{yz-\gamma}$ action is $\mathbb{Z}_2^{yz-f}: f_i\rightarrow f_i^\dagger$, similar to a particle-hole transformation. Now we can consider a four fermion interaction:
\begin{equation}
    H_{int}=V(f_1f_2f_3f_4+h.c.).
\end{equation}
This term is invariant under the $\mathbb{Z}_2$ subsystem symmetries. In addition, it selects a single ground state in the Hilbert space of the 8 zero modes. It is easy to check that the ground state is $|\psi\rangle\sim (1+f_1^\dagger f_2^\dagger f_3^\dagger f_4^\dagger)|0\rangle$, where $|0\rangle$ is the empty state for the $f$ fermions. Under the $\mathbb{Z}_2^{yz-f}$ symmetry, $|0\rangle \rightarrow f_1^\dagger f_2^\dagger f_3^\dagger f_4^\dagger|0\rangle$. Therefore, $|\psi\rangle$ is invariant under this symmetry. 

Since we are able to find a trivial topological defect by turning on symmetric interactions, we can achieve a symmetric gapped bulk by proliferating the topological defects.

\section{A chiral model with $c_-=1$: $\mathbb{Z}^x_3\times \mathbb{Z}^y_3$ subsystem symmetry} 

\subsection{Gapped bulk}
\begin{figure}[h]
    \centering
    \includegraphics[width=2in]{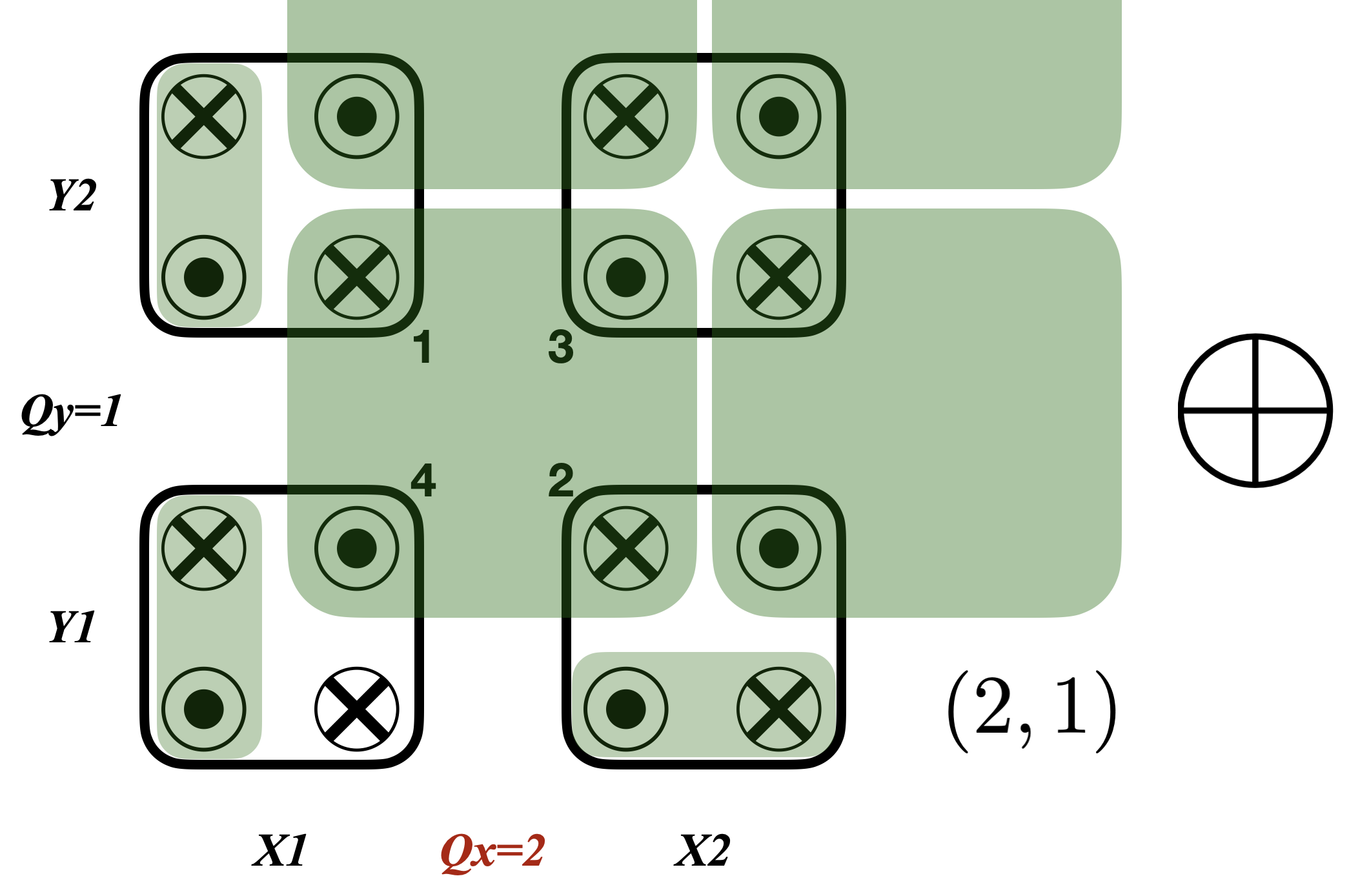}
    \includegraphics[width=2in]{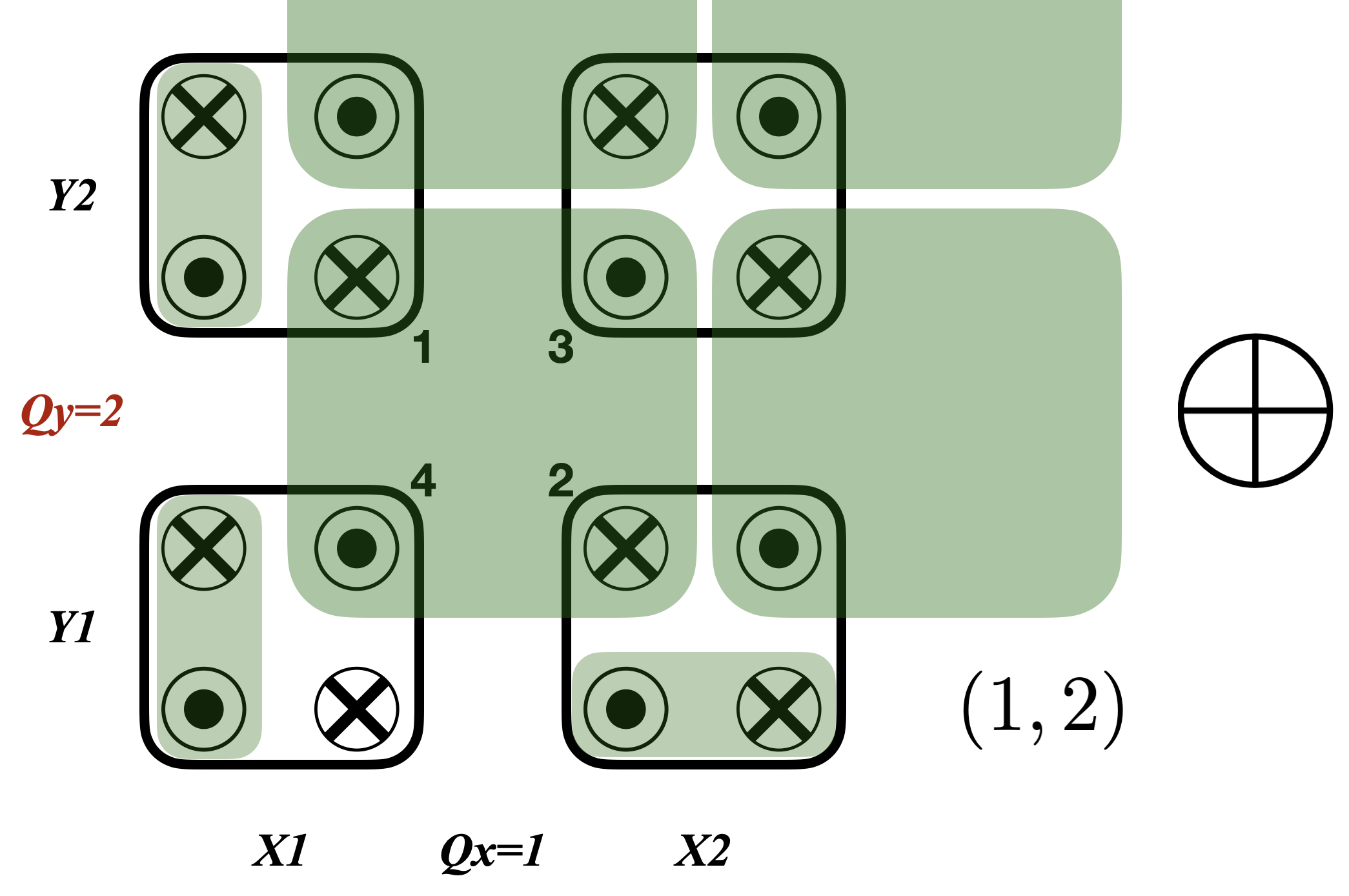}
    \includegraphics[width=2in]{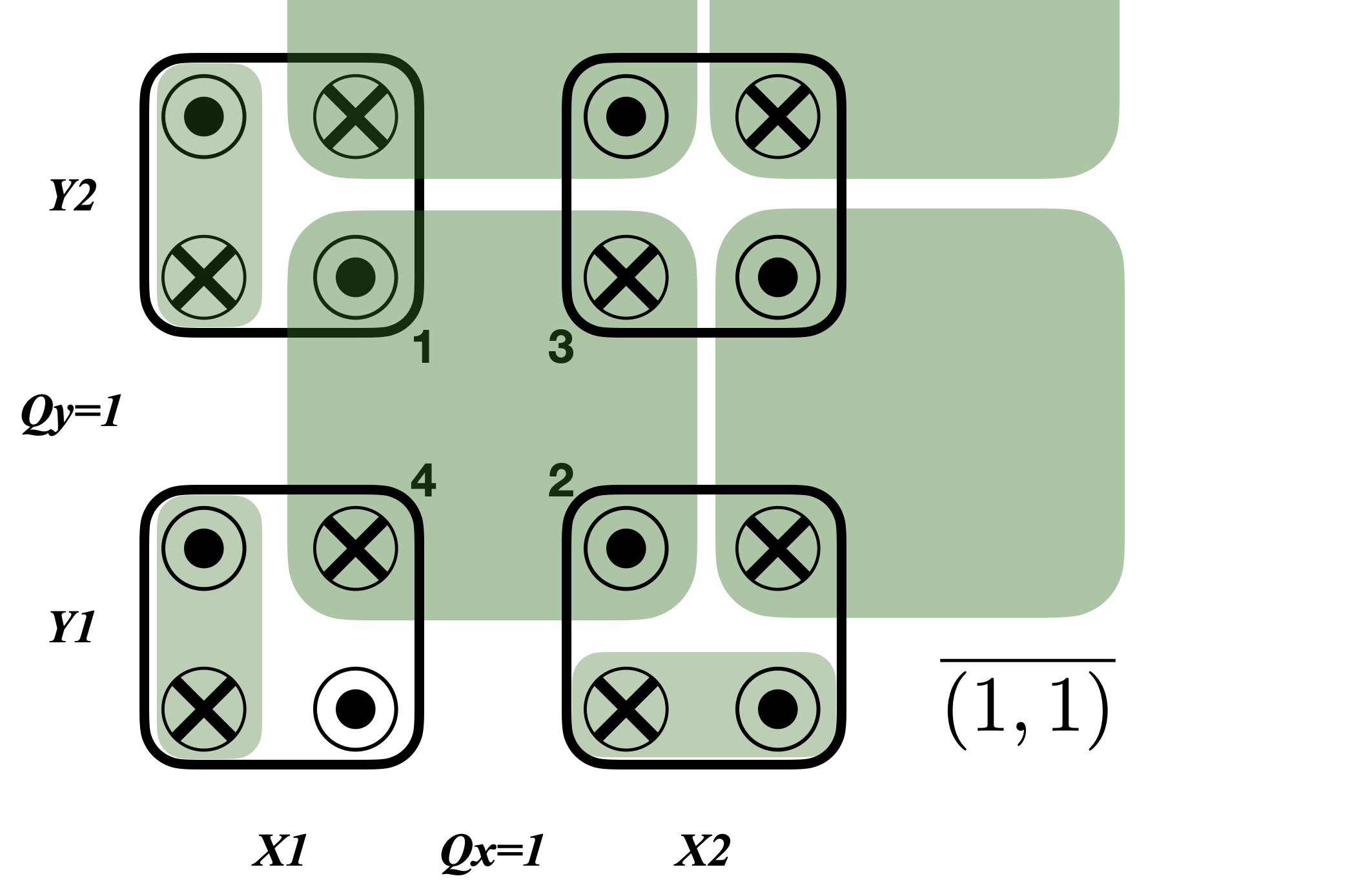}
    \caption{Three copies of the building block. For the first copy, the wires all have $x$-charge 2, $y$-charge 1. The second copy has $x$-charge 1, $y$-charge 2. The third copy has unit charge on $x$ and $y$ directions. However, the chirality of the third copy is reversed.}
    \label{fig2}
\end{figure}

The construction in the main text gives hinge modes with chiral central charge $c_-=N\geq 2$. In order to get smaller chiral central charge, we need to modify our construction. Let us consider a system with $\mathbb{Z}^x_3\times \mathbb{Z}^y_3$ subsystem symmetry. To generate the hinge central charge we want, we need to carefully choose the charge and chirality assignments on the wires. For example, let us consider the structure that is shown in Fig. \ref{fig2} where we make three copies of each bundle of wires. For the first copy, all the wires carry subsystem $x$-charge 2, and subsystem $y$-charge 1. We can conveniently label this copy as $(2,1)$. For the second copy, labeled by $(1,2)$, all the wires carry $y$-charge 2, and $x$-charge 1. Finally, the third copy has $x$-charge 1 and $y$-charge 1, however, the chirality of the 
third copy is reversed. Therefore, we label the third copy as $\overline{(1,1)}$. 

For each wire bundle, there are 4 fermion modes at the intersection between four neighboring unit cells, labeled as $1,2,3,4$ in Fig. \ref{fig2}. We can bosonize these four modes with bosonic variables $\phi_i,i=1,2,3,4$. The $K$ matrix of these four modes is
\begin{equation}
    K=\begin{pmatrix}
1 & 0 & 0 &0\\
0 & 1 & 0 &0\\
0 & 0 & -1 &0\\
0 & 0 & 0 &-1
\end{pmatrix}.
\end{equation}
The charge vectors associated with the subsystem symmetries are given by 
\begin{equation}
    t_{X_1}=\begin{pmatrix}
2\\
0\\
0\\
2
\end{pmatrix},\ \ 
t_{X_2}=\begin{pmatrix}
0\\
2\\
2\\
0
\end{pmatrix}, \ \ 
 t_{Y_1}=\begin{pmatrix}
0\\
1\\
0\\
1
\end{pmatrix},\ \ 
t_{Y_2}=\begin{pmatrix}
1\\
0\\
1\\
0
\end{pmatrix}.
\end{equation}
For a single bundle, it is actually not possible to gap all the modes without breaking any symmetry. The reason is similar to the U(1)$^{xz}\times$U(1)$^{yz}$ case - the 4 fermion modes with these symmetry actions map precisely to the boundary of a 2+1D SPT. However, it is possible to reduce the 4 fermion modes down to 2 bosonic modes by a four fermion interaction 
\begin{equation}
    g_0 \cos(\phi_1+\phi_2-\phi_3-\phi_4).
    \label{intblock}
\end{equation} The two bosonic modes that are left are given by the following vectors,
\begin{equation}
    l_1=\begin{pmatrix}
1\\
0\\
-1\\
0
\end{pmatrix},\ \ 
l_2=\begin{pmatrix}
1\\
0\\
0\\
-1
\end{pmatrix}.
\label{bosons}
\end{equation}
Projecting the $K$-matrix into these two modes, we get an effective $K$-matrix:
\begin{equation}
    K_{eff}=\begin{pmatrix}
0&1\\
1&0
\end{pmatrix},
\label{Keff}
\end{equation}
and effective charge vectors are given by
\begin{equation}
    t_{X_1}=-t_{X_2}=\begin{pmatrix}
2\\
0
\end{pmatrix},\ \ 
 t_{Y_2}=-t_{Y_1}=\begin{pmatrix}
0\\
1
\end{pmatrix}.
\end{equation}

To proceed, for each of our three wire bundle copies we first turn on the above interactions which reduce all of the fermionic modes to a total of six bosonic modes. The three copies together are described by the following effective $K$-matrix:
\begin{equation}
    K_{eff}^{(3)}=\begin{pmatrix}
1&0&0\\
0&1&0\\
0&0&-1
\end{pmatrix}\bigotimes \begin{pmatrix}
0&1\\
1&0
\end{pmatrix},
\label{Keff3}
\end{equation}
with charge vector,
\begin{equation}
t_{X_1}=-t_{X_2}=\begin{pmatrix}
2\\
0\\
1\\
0\\
1\\
0
\end{pmatrix},\ \ 
 t_{Y_2}=-t_{Y_1}=\begin{pmatrix}
0\\
1\\
0\\
2\\
0\\
1
\end{pmatrix}.
\end{equation}

Within this set of modes, we can find three symmetric interaction terms that are bosonic, linearly independent, and mutually commuting so that we can  gap out all the modes without spontaneous symmetry breaking. The explicit interactions are of the form of $\sum_{i=1}^{3}g_i \cos(L_i^T\phi)$ where
\begin{align}
    L_1^T&=(1,0,0,0,1,0); \nonumber\\
    L_2^T&=(0,0,-1,0,1,0); \nonumber\\ L_3^T&=(0,1,0,-1,0,1).
\end{align}
It is easy to verify that if any operator of the form $e^{iL^T\phi}$ commutes with the above interactions, then $L$ must be an integral linear combination of $L_1,L_2$, and $L_3$, which means there is no gapless mode left in the regime where these interactions are strong. 

Beyond the technical considerations, we can provide intuition for why our bundle charges and such interactions can lead to a gapped bulk. This is again due to the $\mathbb{Z}_3$ classification of BSPTs with $\mathbb{Z}_3\times \mathbb{Z}_3$ symmetry. Let us label these BSPTs by an integer index $\nu$, which is well-defined only mod 3. The low-energy bosonic modes in the $(2,1)$, $(1,2)$ and $\overline{(1,1)}$ sectors are equivalent to the edge modes of BSPTs having topological index $2$,$2$, and $-1$ respectively. So the total index is $\nu=2+2-1=3=0$ mod $3$, i.e., it is trivial, which implies that the modes are anomaly free and gappable.  

\subsection{Hinge state}

\begin{figure}[h]
    \centering
    \includegraphics[width=0.4\textwidth]{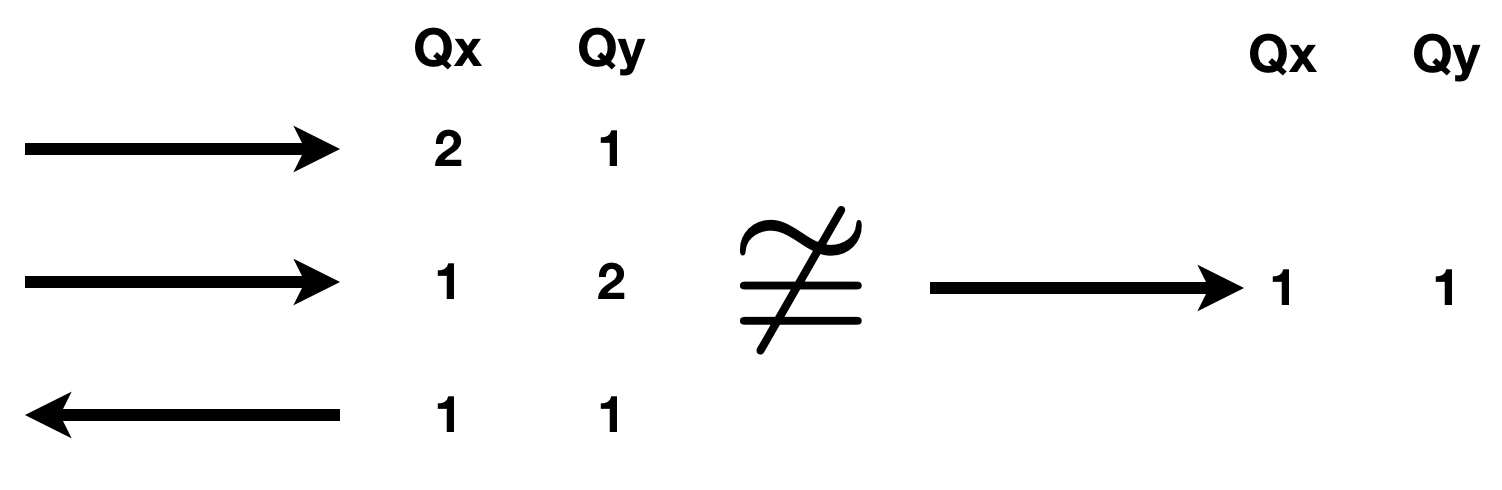}
    \caption{The hinge modes from construction in Fig. \ref{fig2} are NOT equivalent to a single chiral fermion with susbystem $x$-charge 1 and subsystem $y$-charge 1.}
    \label{fig3}
\end{figure}

From Fig. \ref{fig2}, we observe that the hinge has three modes, two left moving and one right moving, and with an interesting charge assignment, summarized in Fig. \ref{fig3}. This set of modes indeed has total chiral central charge $c_-=1$. Unfortunately, these modes are NOT equivalent to a single chiral mode from an anomaly point of view as shown in Fig. \ref{fig3}. The single chiral mode shown on the right of Fig. \ref{fig3} has a mixed anomaly between $x$-symmetry and $y$-symmetry, the mixed anomaly index is $k_{xy}=Q_x\times Q_y=1\times 1=1$ mod 3. However, this anomaly is not matched by the left side of Fig. \ref{fig3}, whose index is $k_{xy}'=2\times 1+1\times 2-1\times 1=0$ mod 3.

Actually, within this construction, it is not possible to realize a hinge mode that is equivalent to a single chiral fermion mode, because the gapping condition of the bulk is equivalent to the vanishing condition of the mixed anomaly between the $x$-symmetry and $y$-symmetry.

\section{Another chiral model with $c_-=1$: $\mathbb{Z}_2^x\times \mathbb{Z}_3^y$ subsystem symmetry}

\begin{figure}[h]
    \centering
    \includegraphics[width=2in]{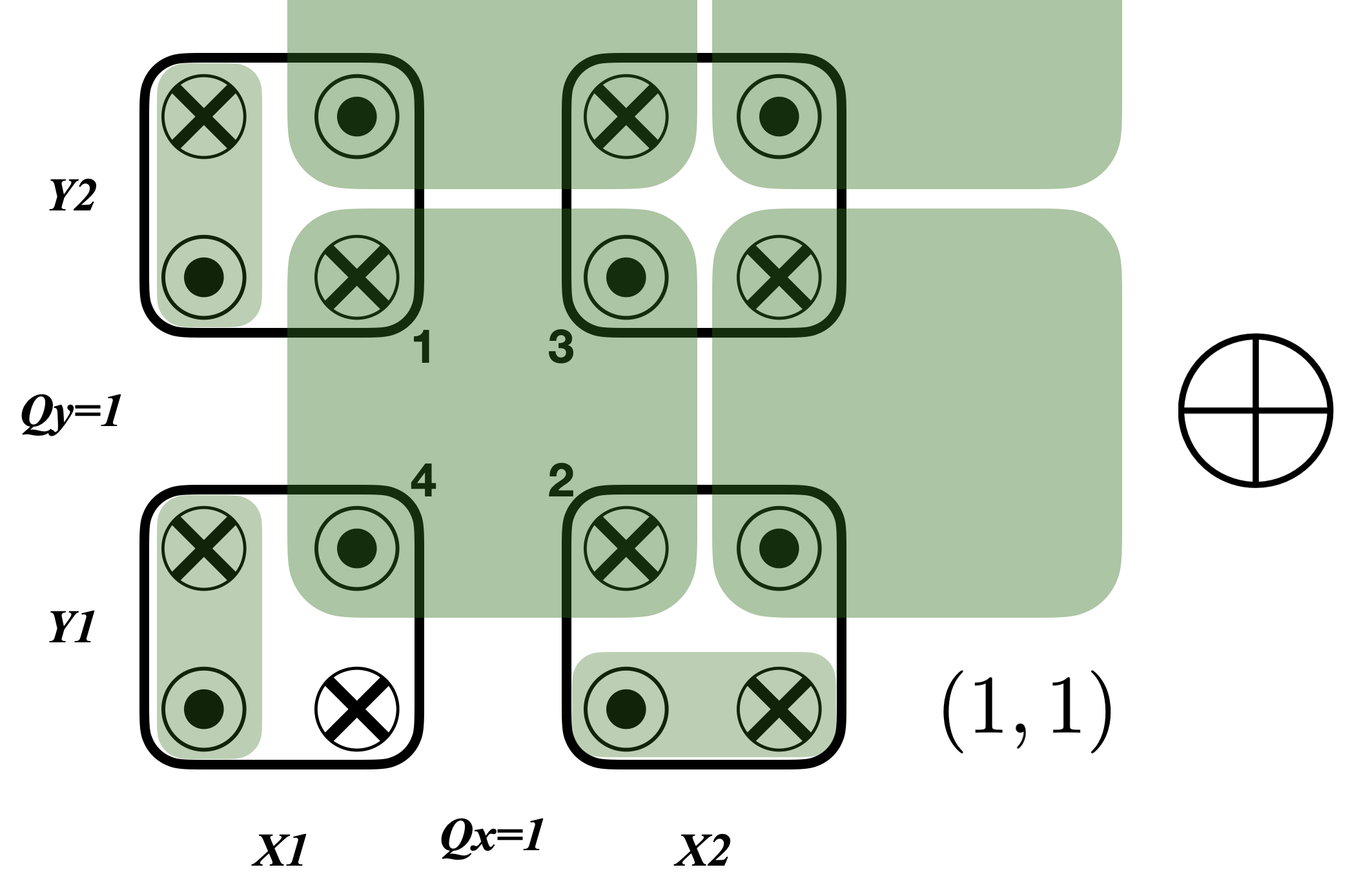}
    \includegraphics[width=2in]{Z2Z311n.png}
    \includegraphics[width=2in]{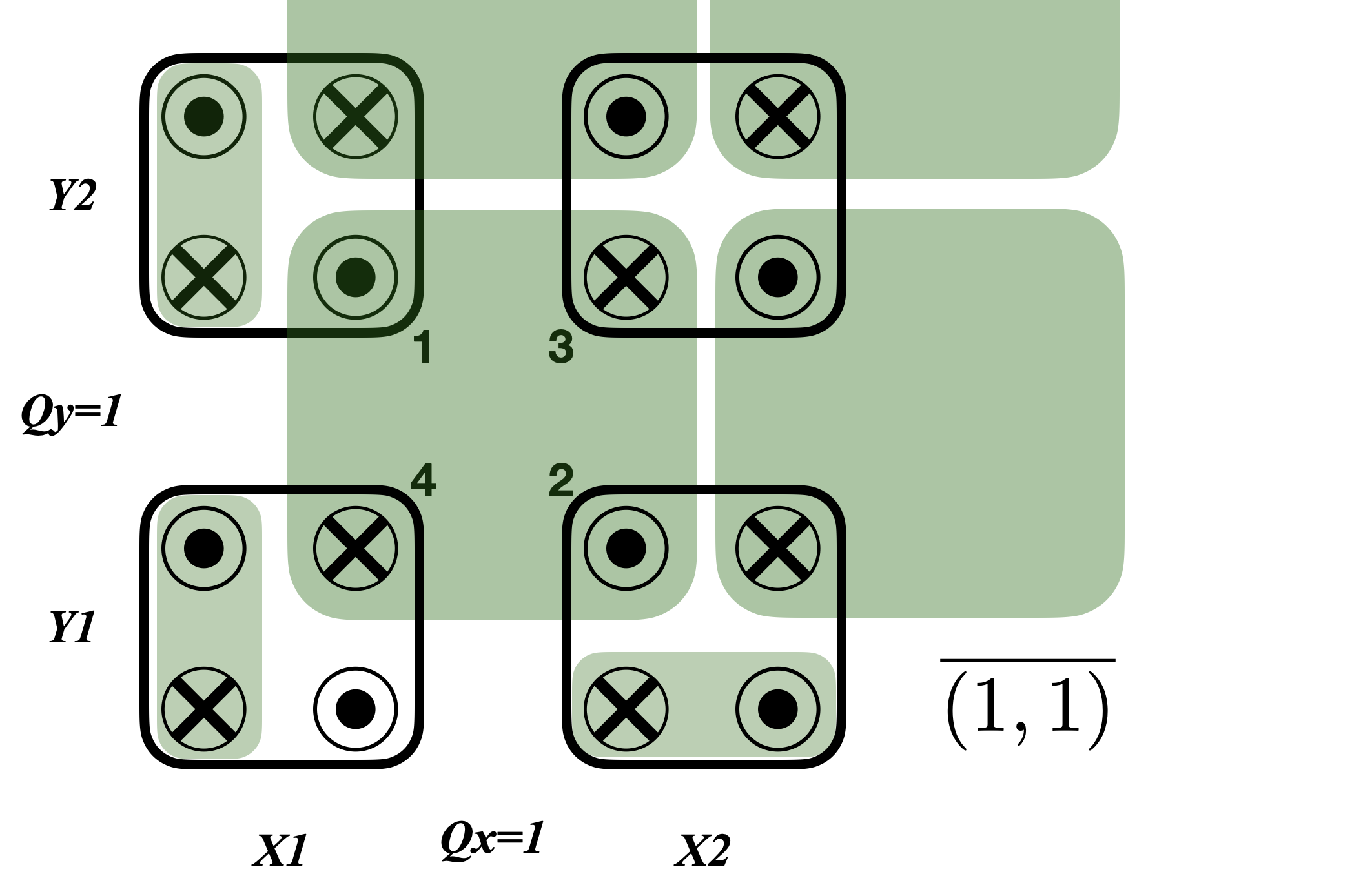}
    \caption{Three copies of the building block. For the first two copies, the wires all have $x$-charge 1, $y$-charge 1. The third copy has unit charge on $x$ and $y$ directions as well, however, the chirality of the whole system is reversed.}
    \label{fig4}
\end{figure}

Next we can consider the case of subsystem symmetry of $\mathbb{Z}_2^x\times \mathbb{Z}_3^y$, this notation means that the subsystem symmetry along the $xz$-plane is $\mathbb{Z}_2$ and the susbsystem symmetry along the $yz$-plane is $\mathbb{Z}_3$. Interestingly, the classification of 2+1D bosonic SPTs having $\mathbb{Z}_2\times \mathbb{Z}_3$ symmetry is trivial. This implies that, if we just take a single wire bundle as in the upper panel of Fig. \ref{fig4} and consider the symmetry to be $\mathbb{Z}_2^x\times \mathbb{Z}_3^y$, we should be able to gap out the bulk without breaking the symmetries. However, it is not the case here, as  we cannot find symmetric terms that gap out the modes for the bundle in the upper panel of Fig. \ref{fig4}  without spontaneously breaking the symmetries. In other words, within the single bundle we cannot construct a trivial gapped bulk even if we break the symmetry down to $\mathbb{Z}_2^x\times \mathbb{Z}_3^y$. This is actually not in contradiction to the classification of 2+1D SPTs. The reason is that the classification of 2+1D SPTs only gives \textit{stable equivalence} of the edge theory\cite{StableEq}. Hence, we expect that the modes of a single building block, although anomaly free, in this case can only be trivialized by mixing with additional trivial modes. 

Nonetheless, we can design an alternative model to accomplish our goal. The alternative model in the notation introduced above is two have three blocks with charge assignments $(1,1)\oplus(1,1)\oplus\overline{(1,1)}$ shown in Fig. \ref{fig4}. For this construction, we can find the appropriate term to symmetrically gap out the bulk, and on the hinge, we can realize a single chiral fermion mode. 

Explicitly, within each bundle, we turn on the interaction in Eq. \ref{intblock} to reduce the 4 fermionic modes into 2 bosonic modes. The effective $K$-matrix is the same as in Eq. \ref{Keff3}. However, the charge vectors are different and given by:
\begin{equation}
t_{X_1}=-t_{X_2}=\begin{pmatrix}
1\\
0\\
1\\
0\\
1\\
0
\end{pmatrix},\ \ 
 t_{Y_2}=-t_{Y_2}=\begin{pmatrix}
0\\
1\\
0\\
1\\
0\\
1
\end{pmatrix}.
\end{equation}
We have to bear in mind that the $x$-charges are defined mod 2 and $y$-charges are defined mod 3. With these constraints, we can find three interaction terms that can gap out all the modes. The $L$ vectors corresponding to the three terms are given by 
\begin{align}
    L_1^T&=(1,0,-1,0,0,0); \nonumber\\ L_2^T&=(1,0,0,0,1,0); \nonumber\\
    L_3^T&=(0,1,0,1,0,1).
\end{align}
As discussed in the previous Appendix, these $L$ vectors generate cosine  interactions, and for our choice of $L$ vectors, the interactions will leave the bulk symmetric, gapped, and non-degenerate. 

\begin{figure}[h]
    \centering
    \includegraphics[width=0.4\textwidth]{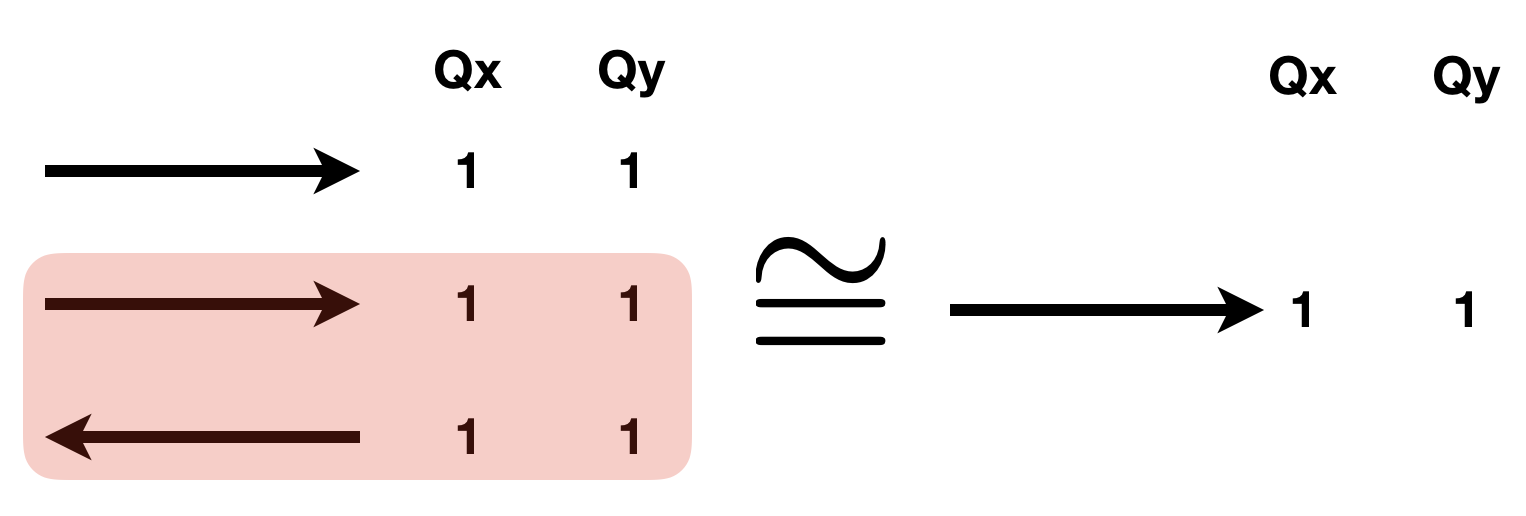}
    \caption{The hinge modes from construction in Fig. \ref{fig4}. We can safely gap out a pair of counter-propagating modes with same symmetry charges and arrive a hinge mode which only contain a single chiral fermion.}
    \label{fig5}
\end{figure}

The hinge modes for this model are very simple, as shown in Fig. \ref{fig5}. Indeed, for this model, we can realize a single chiral fermionic mode on the hinge by turning on a intra-unit-cell tunneling to gap out a pair of helical modes. 

An interesting question is what is the classification of the HOSPT that we just constructed. We argue that the classification should be $\mathbb{Z}_2$. Namely, for two copies of the HOSPT, it is possible to gap out the hinge modes by attaching subsystem symmetric 2D layers on the surface. To see this, let us first consider the surface that is normal to the $x$-direction. On such a surface, the subsystem $\mathbb{Z}_2$ symmetry implies that for each unit cell along the $z$-direction, the fermion parity is separately conserved. The subsystem $\mathbb{Z}_3$ symmetry becomes a global $\mathbb{Z}_3$ symmetry of the entire surface layer. A pure 2D system with such a symmetry assignment will always have chiral central charge $c=4N, N\in \mathbb{Z}$, on its boundary along the $z$-direction, see the next section for an argument. This indicates that, if we have 4 copies of the constructed HOSPT, the hinge modes can be gapped out by attaching a pure 2D subsystem symmetric surface state and turning on symmetric tunneling terms. Therefore, the classification of the HOTSC state is at most $\mathbb{Z}_4$. Now let us consider the $yz$ surface. The subsystem $\mathbb{Z}_3$ symmetry becomes a subsystem $\mathbb{Z}_3$ symmetry for the surface, namely each unit cell along $z$-direction has an individual $\mathbb{Z}_3$ symmetry. The subsystem $\mathbb{Z}_2$ symmetry becomes the fermion parity for the entire surface. Running a similar argument as the next section, we find that a pure 2D system with such subsystem symmetry assignment requires the boundary central charge to be $c=6M$, $M\in\mathbb{Z}$. Therefore, for 6 copies of the HOTSC, the hinge modes can be gapped out by attaching a pure 2D system on the $yz$ surface. Combining the information from both the $xz$ and $yz$ surface, the classification of the HOTSC is $\mathbb{Z}_2$. 

\section{2D Topological superconductor protected by subsystem symmetry}\label{App:2DTSC}
In this section, we will consider a 2D topological superconductor(TSC) protected by subsystem symmetry where the fermion parity charge is conserved on all vertical rows (parallel to the $y$-direction). We will show that these subsystem symmetric TSCs must have $8N$ chiral Majorana edge modes, i.e., they have an edge chiral central charge of $c=4N$. We will do this in two steps. First, we explicitly construct a subsystem symmetric TSC with $8$ Majorana edge modes. Second, we will argue that any subsystem symmetric TSC where the Majorana edge modes do not come in multiples of 8 is inconsistent. 

\begin{figure}[h]
    \centering
    \includegraphics[width=0.3\textwidth]{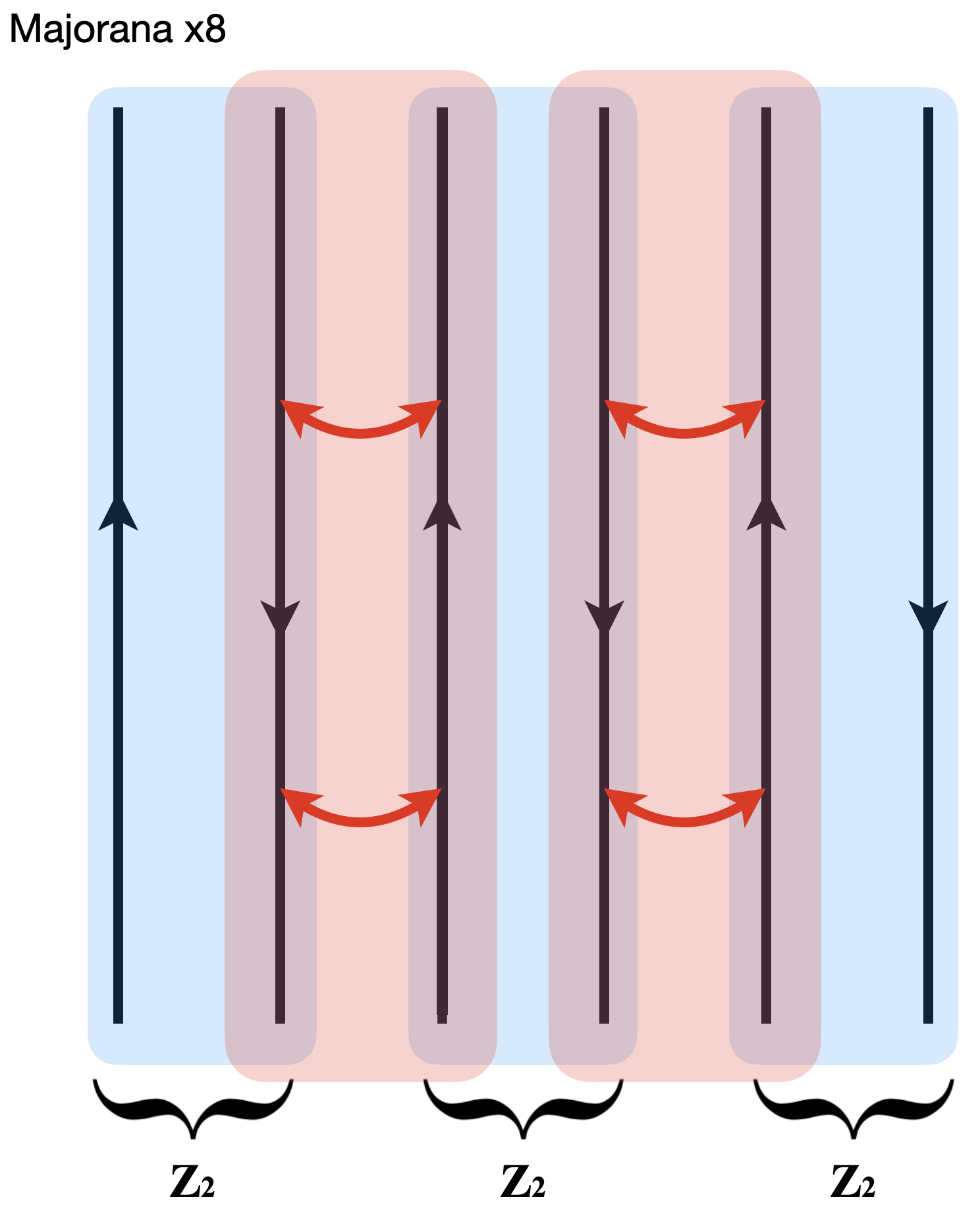}
    \caption{Coupled wire construction of a 2D TSC. Each blue block is a unit cell. The fermion parity of each unit cell is individually conserved.}
    \label{2d}
\end{figure}

We will build a subsystem symmetric TSC via the coupled wire construction. As shown in Fig.~\ref{2d} we couple 1D wires along the vertical direction that each contain eight helical Majorana modes. Our aim is to couple the wires in a way that preserves the subsystem fermion parity symmetry, and gaps out the bulk while leaving chiral Majorana modes on the edges. To that end, we couple the eight left-moving chiral Majoranas from $i$th wire to the eight right-moving chiral Majoranas at the adjacent $(i+1)$th wire. The fermion parity is conserved on each vertical row, therefore, single particle inter-wire tunnelings are forbidden. Instead, the symmetry requires inter-wire tunneling in terms of Majorana pairs. Hence, we couple the wires via a quartet Majorana interaction that preserves subsystem fermion parity symmetry (denoted by the red shading in Fig.~\ref{2d}). 

Explicitly, each coupled wire unit contains eight helical Majorana modes:
\begin{align}
    &H=\eta^T(k_z \sigma^{300})\eta \\
    & \mathbb{Z}_2^{i}: \eta \rightarrow -\sigma^{300} \eta \nonumber \\
    & \mathbb{Z}_2^{i+1}: \eta \rightarrow \sigma^{300} \eta,
    \label{interwire2}
\end{align}
where the eight left and right moving modes carry independent fermion parity symmetries $\mathbb{Z}_2^{i}$ and $\mathbb{Z}_2^{i+1}$. The eight copies of helical Majorana modes and the symmetry action above precisely map to the edge of 8 copies of a bilayer consisting of one $p+ip$ and one $p-ip$ superconductor with individual fermion parity for the $\pm$ layers. The interacting classification for $p\pm ip$ with individual fermion parities for the two chiralities is $\mathbb{Z}_8$, which means the system defined in Eq. \ref{interwire2} is anomaly free. Indeed, we can just adapt the method we discussed in Sec.~\ref{sec:HOTSC} and add the four fermion interaction term as in Eq.~\ref{eq:LatticeHamiltonianHOTSC} to gap out the coupled helical wires in the bulk. The resultant state has a symmetric gapped bulk, and gapless edge modes that contain eight chiral Majoranas whose total central charge is $c=4$. Such a subsystem symmetry protected TSC has a $\mathbb{Z}$ classification, since the boundary state is chiral and exhibits a gravitational anomaly. Notably, the central charge for this class of 2D TSCs has to be an integer of four, namely $c=4\mathbb{Z}$, otherwise it is not possible to construct a gapped bulk.

Having shown that there exist 2D subsystem symmetric TSC with $8N$ Majorana edge modes, we will now show that a subsystem symmetric TSC with any other number of Majorana edge modes is inconsistent. To do this, let us consider a subsystem symmetric TSC with $N'$ Majorana edge modes where $N'$ is \textit{not} a multiple of $8$. We can imagine cutting such a system between two vertical rows. This will lead to two edges, one with $N'$ right moving Majorana modes and one with $N'$ left moving modes. Due to the subsystem symmetry, the parity of each edge is separately conserved. We can now consider reversing the cutting procedure to symmetrically glue the edges back together. This will involve gapping out the $N'$ right moving and $N'$ left moving Majorana fermions, while preserving the parity of both the left and right movers separately. However, it has been shown that such a gapping procedure is possible only when $N'$ is a multiple of $8$\cite{fidkowski2010effects,bentov2015fermion}. Hence, we can conclude that a subsystem symmetric TSC must necessarily have $8N$ chiral Majorana edge modes.

\section{Ground State Degeneracy of the FHOTI}\label{App:GSD}
In the main text, the FHOTI was defined in terms of the bosons: 
\begin{equation}\begin{split}
&\tilde{\phi}^i_{R\sigma} = \frac{n+1}{m}\phi^i_{R \sigma} + \frac{n}{m}\phi^i_{L \sigma},\\
&\tilde{\phi}^i_{L\sigma} = \frac{n+1}{m}\phi^i_{L \sigma} + \frac{n}{m}\phi^i_{R \sigma}, 
\label{Aeq:FHOTfracBosons}\end{split}\end{equation}
where $\phi^i_{R \sigma}$ are compact bosons satisfying $\phi^i_{R/L \sigma} \equiv \phi^i_{R/L \sigma} + 2\pi$ for each chirality, spin, and flavor index $i$. This indicates that the bosonic fields in Eq. \ref{Aeq:FHOTfracBosons} satisfy $(\tilde{\phi}^i_{R \sigma}, \tilde{\phi}^i_{R \sigma}) \equiv (\tilde{\phi}^i_{R \sigma} + \pi/m, \tilde{\phi}^i_{R \sigma}-\pi/m)$. 

To determine the ground state degeneracy of the FHOTI, we need to determine operators that commute with the interactions in Eq. \ref{eq:FHOTIintraBoson} and \ref{eq:FHOTIinterBoson}, and are invariant under the equivalence relationships of the compact bosons in Eq. \ref{Aeq:FHOTfracBosons}. These constraints are satisfied by the following operators,
\begin{equation}\begin{split}
&G_{x1}(n^0_x) = \exp(i \sum_{ n_y} [\tilde{\phi}^1_{ \downarrow, \bm{r}} + \tilde{\phi}^4_{ \downarrow, \bm{r}}  + \tilde{\phi}^2_{ \downarrow, \bm{r}'''} + \tilde{\phi}^3_{ \downarrow, \bm{r}'''}])\\
&G_{x2}(n^0_x) = \exp(i \sum_{ n_y} \sum_i [\tilde{\phi}^i_{ \uparrow, \bm{r}} - \tilde{\phi}^i_{ \downarrow, \bm{r}}])\\
&G_{y1}(n^0_y) = \exp(i \sum_{ n_x} [\tilde{\phi}^1_{ \uparrow, \bm{r}} + \tilde{\phi}^3_{ \uparrow, \bm{r}}  + \tilde{\phi}^2_{ \uparrow, \bm{r}'''} + \tilde{\phi}^4_{ \uparrow, \bm{r}'''}])\\
&G_{y2}(n^0_y) = \exp(i \sum_{ n_y} \sum_i [\tilde{\phi}^i_{ \uparrow, \bm{r}} - \tilde{\phi}^i_{ \downarrow, \bm{r}}]),\\
\label{Aeq:StringOperators}\end{split}\end{equation}
where $\tilde{\phi}^i_{ \sigma, \bm{r}} = \tilde{\phi}^i_{R \sigma, \bm{r}} + \tilde{\phi}^i_{L \sigma, \bm{r}}$. For $G_{x1}(n^0_x)$, and $G_{x2}(n^0_x)$ the coordinate $\bm{r} = (n^0_x \hat{x}, n_y \hat{x})$, and for $G_{y1}(n^0_y)$, and $G_{y2}(n^0_y)$ the coordinate $\bm{r} = (n_x \hat{x}, n^0_y \hat{x})$. These four operators all commute with the interactions in Eq. \ref{eq:FHOTIintraBoson} and \ref{eq:FHOTIinterBoson}, and are invariant under the equivalence relationship of the bosons in Eq. \ref{Aeq:FHOTfracBosons} (which leaves $\tilde{\phi}^i_{\sigma}$ invariant). Clearly, these operators are non-local, and scale linearly with the length of the system.

Based on the interactions in Eq. \ref{eq:FHOTIintraBoson} and \ref{eq:FHOTIinterBoson} these operators all take values $\exp(i 2\pi m' / m)$ for integer $m'$, $0\leq m' < m$. These operators are not all unique, as
\begin{equation}\begin{split}
\prod_{n^0_x} G_{x2}(n^0_x) = \prod_{n^0_y} G_{y2}(n^0_y) = \prod_{n^0_x,n^0_y}G_{x1}(n^0_x) G_{y1}(n^0_y).
\end{split}\end{equation}
Hence, for a system size $L_x \times L_y \times L_z$, there are $2L_x + 2L_y -2$ unique operators that commute with the interactions, and are invariant under the boson equivalence relationships. Since each of these operators can take on $m$ different values, there are a total of $m^{2L_x+2L_y -2}$ different ground states.

\end{document}